\documentclass[journal]{IEEEtran}
\usepackage{amsmath,amsfonts}
\usepackage{algorithmic}
\usepackage{algorithm}
\usepackage{array}
\usepackage[caption=false,font=normalsize,labelfont=sf,textfont=sf]{subfig}
\usepackage{textcomp}
\usepackage{stfloats}
\usepackage{url}
\usepackage{verbatim}
\usepackage{graphicx}
\usepackage{cite}
\usepackage{tabularx}
\usepackage{multirow}
\usepackage{xcolor}
\usepackage{balance}
\usepackage[bookmarks=false]{hyperref}
\begin{document}

\title{Artificial Intelligence for the Metaverse: A Survey}

\author{Thien Huynh-The,~\IEEEmembership{Member,~IEEE,} Quoc-Viet Pham,~\IEEEmembership{Member,~IEEE,} Xuan-Qui Pham,\\ 
Thanh Thi Nguyen, Zhu Han,~\IEEEmembership{Fellow,~IEEE,} and Dong-Seong Kim,~\IEEEmembership{Senior,~IEEE}
\thanks{Thien Huynh-Them, Xuan-Qui Pham, and Dong-Seong Kim are with the Department of IT Convergence, Kumoh National Institute of Technology, Gumi, Gyeongsangbuk-do 39177, Republic of Korea (email: thienht@kumoh.ac.kr, pxuanqui@kumoh.ac.kr, dskim@kumoh.ac.kr).}
\thanks{Quoc-Viet Pham is with the Korean Southeast Center for the 4th Industrial Revolution Leader Education, Pusan National University, Busan 46241, Republic of Korea (email: vietpq@pusan.ac.kr).}
\thanks{Thanh Thi Nguyen is with the School of Information Technology, Deakin University, Waurn Ponds, VIC 3216, Australia (email: thanh.nguyen@deakin.edu.au).}  
\thanks{Zhu Han is with the Department of Electrical and Computer Engineering, University of Houston, Houston, TX 77004 USA (e-mail: zhan2@uh.edu).}}



\maketitle

\begin{abstract}

Along with the massive growth of the Internet from the 1990s until now, various innovative technologies have been created to bring users breathtaking experiences with more virtual interactions in cyberspace. Many virtual environments with thousands of services and applications, from social networks to virtual gaming worlds, have been developed with immersive experience and digital transformation, but most are incoherent instead of being integrated into a platform. In this context, metaverse, a term formed by combining meta and universe, has been introduced as a shared virtual world that is fueled by many emerging technologies, such as fifth-generation networks and beyond, virtual reality, and artificial intelligence (AI). Among such technologies, AI has shown the great importance of processing big data to enhance immersive experience and enable human-like intelligence of virtual agents. In this survey, we make a beneficial effort to explore the role of AI in the foundation and development of the metaverse. We first deliver a preliminary of AI, including machine learning algorithms and deep learning architectures, and its role in the metaverse. We then convey a comprehensive investigation of AI-based methods concerning six technical aspects that have potentials for the metaverse: natural language processing, machine vision, blockchain, networking, digital twin, and neural interface, and being potential for the metaverse. Subsequently, several AI-aided applications, such as healthcare, manufacturing, smart cities, and gaming, are studied to be deployed in the virtual worlds. Finally, we conclude the key contribution of this survey and open some future research directions in AI for the metaverse.

\end{abstract}

\begin{IEEEkeywords}
3D virtual world, artificial intelligence, deep learning, machine learning, metaverse.
\end{IEEEkeywords}

\section{Introduction}
\begin{figure}[!t]
	\centering
	\includegraphics[width=1.0\linewidth]{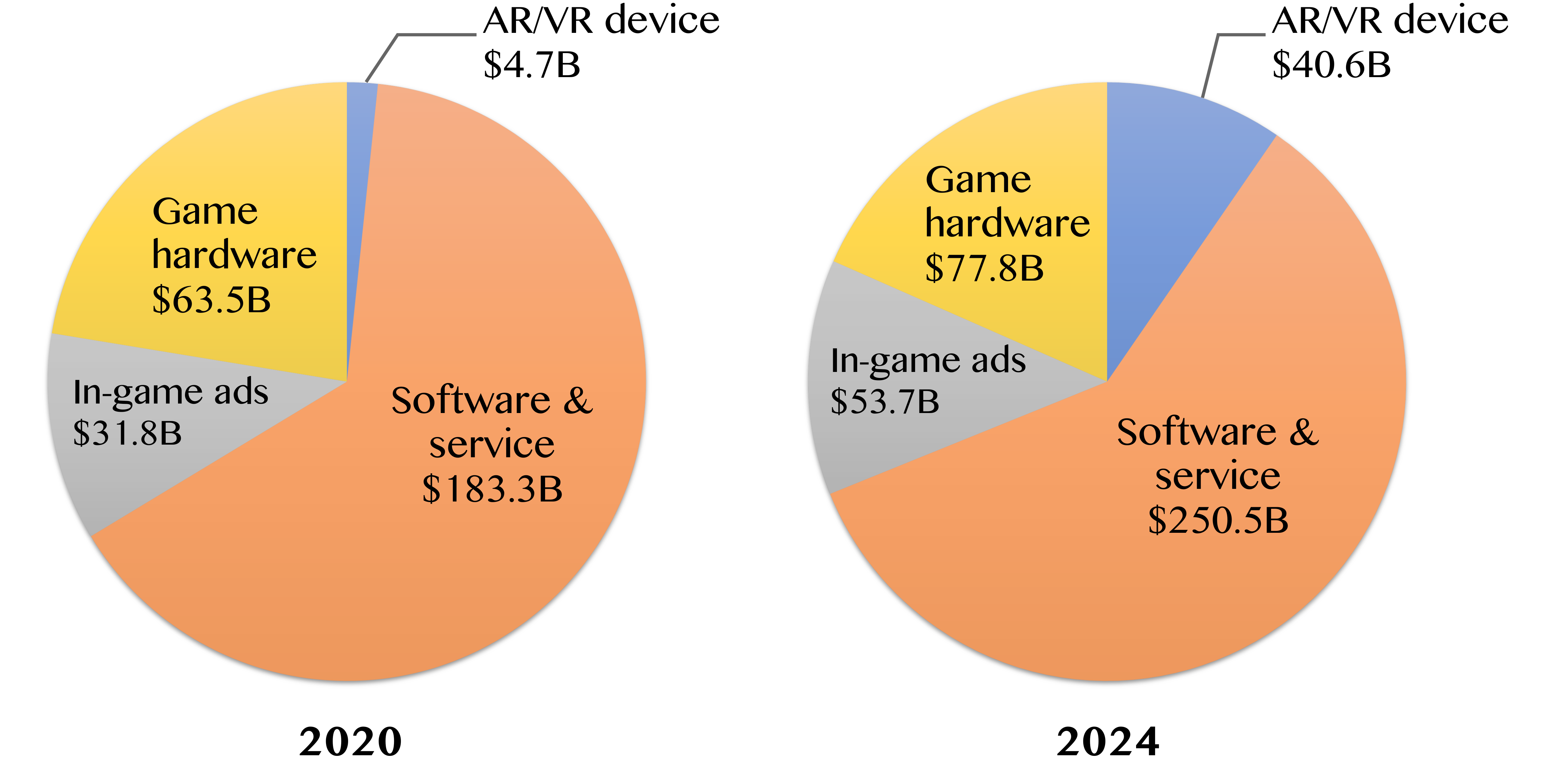}
	\caption{Gaming revenue growth aided by 3D virtual worlds.}
	\label{fig_market}
\end{figure}

\IEEEPARstart{S}{ince} Facebook rebranded itself as Meta, announced by Mark Zuckerberg in October 2021, the marvelous concept regarding the new name has become a hot trend on social media and received huge attention and much more discussions by various communities, including academia and industry. 
Besides Meta, some big tech companies have some metaverse investment and development activities, such as Microsoft bought Activision Blizzard, a video game holding company, for $\$68.7$ billion as the deal of gaming expansion into the metaverse.
Recently, Metaverse Group, a metaverse real estate investment company bought a parcel of land on a decentralized virtual reality platform known as Decentraland for a shocking price $\$2.43$ million, and recorded as the highest ever amount for a virtual real estate. 
A famous rapper who bought a plot of land in the Sandbox metaverse for $\$450,000$ is Snoop Dogg, in which this rapper can hold virtual events like music festivals and concerts to bring an immersive experience to the audience participating in the virtual world via the virtual reality technology.
In the near future, the metaverse is realized as the next big technology and currently attracting online game makers, internet finance businesses, social networks, and other technology leaders. 
The Seoul metropolitan government just very recently announced a plan called Metaverse Seoul that creates a virtual communication ecosystem for all municipal administrative areas, such as culture, tourism, economic, educational, and civic service.
Besides providing different business support services and facilities, the Metaverse Seoul will offer some specialized services for people with disabilities to take pleasure in safety and convenience contents using extended reality (XR) technology.
Based on the analysis of Bloomberg Intelligence~\cite{kanterman2021metaverse}, the global metaverse revenue opportunity will increase from USD $500$ billion in 2020 to USD $800$ billion in 2024, in which the online game industry will take half of the global revenue. Remarkably, the video game companies and studios have some plans to upgrade existing traditional games to three-dimensional (3D) virtual world convolving social networks, in which some attractive activities, such as live entertainment and media advertising events, can be held besides gaming.
In Fig.~\ref{fig_market}, the revenue of virtual reality (VR) hardware and in-game advertisement significantly increases through the advancement of virtual activities in the metaverse.

\begin{figure*}[!t]
	\centering
	\includegraphics[width=170mm]{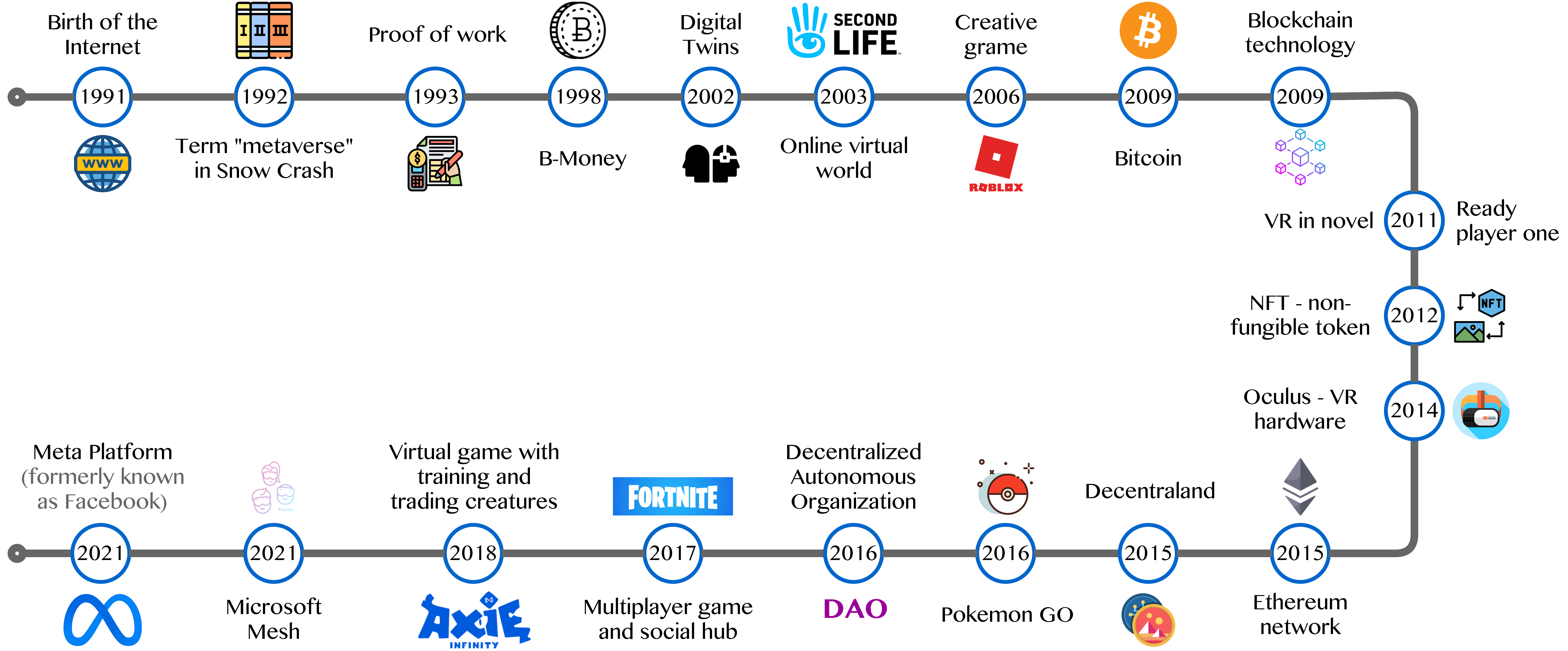}
	\caption{A timeline of the metaverse development involving primary events from 1991 to 2021.}
	\label{fig_timeline}
\end{figure*}

\begin{figure}[!t]
	\centering
	\includegraphics[width=60mm]{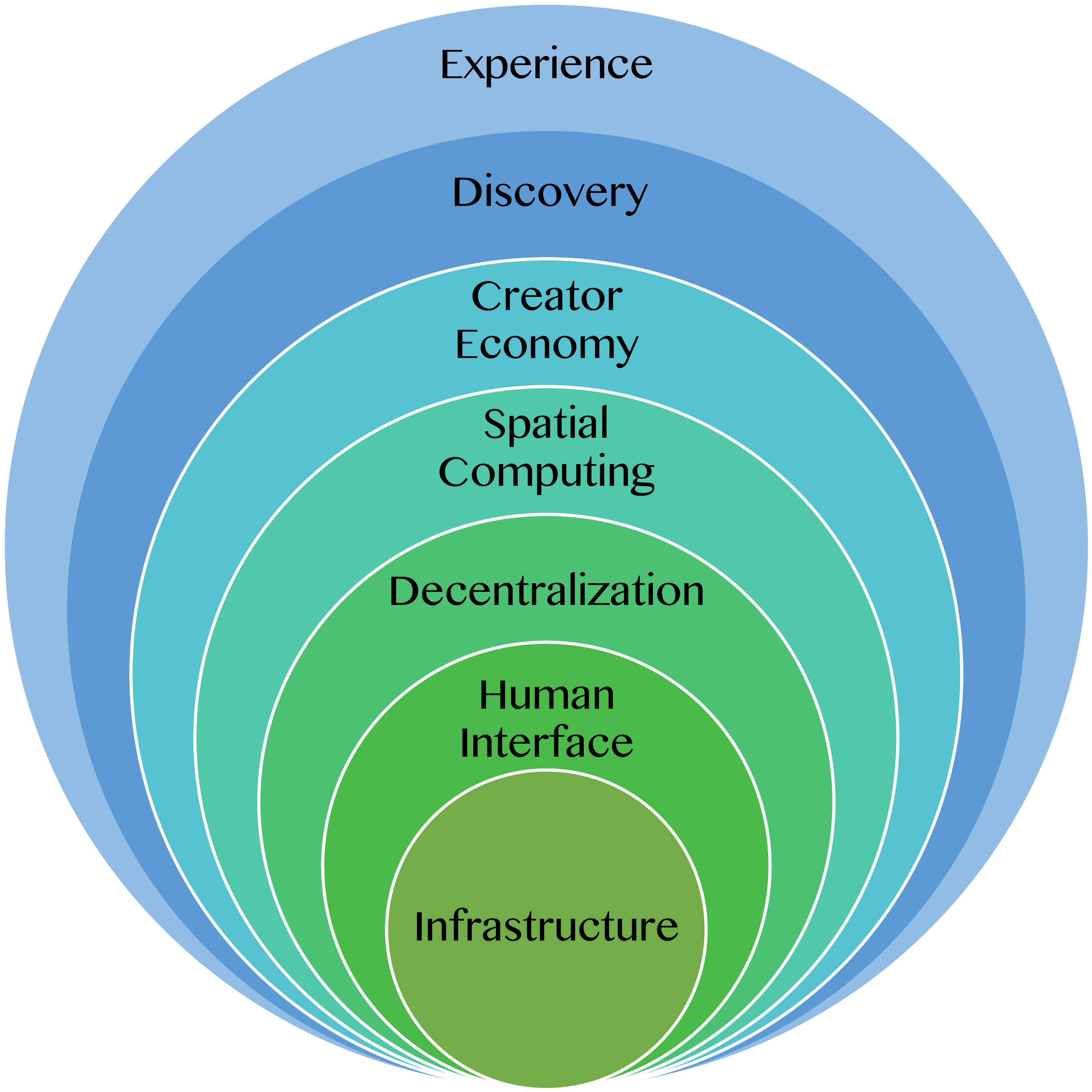}
	\caption{Seven layers of a metaverse platform.}
	\label{fig_metalayer}
\end{figure}

The metaverse is not a new idea because it has circulated along with the development of the Internet and other technologies for decades.
Fig.~\ref{fig_timeline} describes the timeline of the metaverse development that involves many primary events, from the birth of the Internet and the first mention in literature to the first virtual world project with Second Life and recent metaverse projects of big tech companies like Microsoft and Facebook.
Metaverse is the term formed by combining Meta and Universe~\cite{WikipediaMetaverse}, which may be first mentioned in the dystopian cyberpunk novel Snow Crash in 1992 to describe a virtual reality world called the matrix. At present, the metaverse is defined as a shared virtual 3D world or even multiple cross-platform worlds that can provide users a comprehensively immersive experience with interactive and collaborative activities.
Besides virtual places and constructions fixed in the virtual world, many other entities, such as objects, user identities, and digital goods, can be exchanged between different virtual worlds and even reflected into the reality world~\cite{park2022metaserve}.
Few recent years have witnessed an unprecedented explosion of the metaverse, mostly derived from 3D gaming, which is fueled by the improvement of hardware (e.g., big data storage infrastructure, wireless communication networks, built-in sensors, and graphic processing unit - GPU) and the optimization of software (e.g., resource allocation in communications, language processing, and computer vision) to build the virtual world more solidly and creatively.
Different from the traditional metaverse modality that limits immersive experience poorly by insufficient data, the new one not only generates a huge new source of user and behavioral data for enterprises (where users freely make creative content) but also presents a plentiful foundation to deploy artificial intelligence (AI) into various domains, such as natural language processing, computer vision, and neural interface.
Besides, a standard platform built for a modern metaverse should satisfy the following characteristics: virtual world, persistency, scalability, always-on with synchronicity, financial allowance, decentralization, security, and interoperability.
In~\cite{radoff2021themetaverse}, a metaverse platform can include several layers (see Fig.~\ref{fig_metalayer}) which are expressed as follows:
\begin{itemize}
	\item \textit{Infrastructure}: 5G, 6G, WiFi, cloud, data center, central processing units, and GPUs.
	\item \textit{Human interface}: mobile, smartwatch, smartglasses, wearable devices, head-mounted display, gestures, voice, and electrode bundle.
	\item \textit{Decentralization}: edge computing, AI agents, blockchain, and microservices.
	\item \textit{Spatial computing}: 3D engines, VR, augmented reality (AR), XR, geospatial mapping, and multitasking.
	\item \textit{Creator economy}: design tools, asset markets, E-commerce, and workflow.
	\item \textit{Discovery}: advertising networks, virtual stores, social curation, ratings, avatar, and chatbot.
	\item \textit{Experience}: games, social, E-sports, shopping, festivals, events, learning, and working. 
\end{itemize}

It is not hard to find out the presence of AI inside layers, with machine learning (ML) algorithms and deep learning (DL) architectures, along with their importance in many diversified aspects. 
For instance, many ML algorithms with supervised and unsupervised learning were applied in classification and regression models for voice recognition and other language processing tasks that enable system agents to understand user commands.
With the input data of sensor-based signals collected by multiple devices, such as mobile, smartwatch, and other wearable devices, the complex patterns of human actions can be analyzed and learned for some applications like physical activity recognition that allows a system to perceive user activities and interactions in the virtual world.
Recently, DL has emerged as a powerful AI tool to deal with the practical issue of understanding complicated patterns from large-messy-confusing data.
With considerable success in the computer vision domain, DL is now being leveraged in different domains, such as wireless communications, human-computer interaction, gaming, and finance. 
A few years ago, NVIDIA introduced DL super sampling (DLSS), a groundbreaking technology that exploits the power of DL and other AI algorithms to boost the frame rate while maintaining beautiful and sharp in-game images, thus being potential to improve the visual experience in the metaverse. 
AI was also leveraged to improve game balance in several online multiplayer games by training supervised learning models iteratively until satisfying designers and play-testers.
To dive into a new era of 3D design simulation and collaboration for creating an impressive virtual-reality world, in the metaverse, as rich as the real world, NVIDIA introduced Omniverse\footnote{https://www.nvidia.com/en-us/omniverse/}, an open-and-extensible platform, owning many valuable features, including physically-accurate simulation, multi-user design collaboration, photorealistic and real-time rendering, and AI-accelerated workflows.

\subsection{Our Contributions}
The metaverse platform is built by merging many advanced technologies to bring a completely 3D immersive experience to users, where they can truly interact and collaborate with others in the virtual worlds. Among such technologies like blockchain, XR/VR, and 5G, AI has a silent but important role in the foundation and development of the metaverse. However, understanding how AI can affect and contribute to the metaverse in the technical and application aspects is dubious, especially in the context of which it is neither mentioned in a fancy way like XR/VR nor discussed gloriously on social media like blockchain. No existing work is done to provide a comprehensive review of the role and use of AI in the metaverse. 

In the paper, we convey a comprehensive survey of the existing AI-based works in the technical and application perspectives and further discuss their potentials for the metaverse. In a nutshell, the main contributions of this paper are summarized as follows.
\begin{itemize}
	\item We briefly review AI techniques, including conventional ML algorithms and innovative DL architectures, with various learning strategies like supervised learning, unsupervised learning, and reinforcement learning. Based on that, the role of AI in the metaverse is initially revealed. 
	\item We survey the state-of-the-art AI-powered approaches in six technical aspects, including natural language processing, machine vision, blockchain, networking, digital twin (DT), and neural interface, which show great potential for the metaverse platform.
	\item We investigate the existing AI-aided methods relying on several application aspects, such as healthcare, gaming, manufacturing, smart cities, E-commerce, real estate, and decentralized finance, which receive more interest to be deployed in the virtual world.
	\item We introduce several interesting metaverse projects which have applied AI to enhance the immersive experience and develop user-oriented services. Moreover, some future research directions in AI for the metaverse are discussed.
\end{itemize}

\subsection{Paper Organization}
The remaining of this paper is organized as follows. The fundamentals of blockchain, metaverse, and the role of blockchain in the metaverse are presented in Section~\ref{sec_preliminaries}. 
The adoptions of AI for the metaverse in technical aspects, such as natural language processing, machine vision, blockchain, networking, DT, and neural interface, are investigated in Section~\ref{sec_Technical}. 
In Section~\ref{sec_Application}, we discuss several AI-powered applications to be promisingly developed in the metaverse, including healthcare, manufacturing, smart cities, and gaming, besides other minor areas as E-commerce, human resources, real estate, and decentralized finance.
Some remarkable metaverse projects are presented in Section~\ref{sec_Project}.
Finally, we conclude the paper with some future research directions for the development of the metaverse in Section~\ref{sec_Conclusion}.

\section{AI for The Metaverse: Preliminaries}
\label{sec_preliminaries}
This section briefly conveys a wide spectrum of AI, from traditional ML algorithms to advanced DL networks that embrace different learning mechanisms, and then articulates the role of AI in the metaverse.

\subsection{Categorization of AI}
This part reviews some common AI/ML algorithms that are potential for the metaverse.
Fundamentally, most of the existing AI/ML algorithms can be categorized into two sectors: conventional techniques and advanced techniques, which are studied for three principal problems: clustering, classification, and regression.

\subsubsection{Conventional Techniques}
Conventional AI/ML algorithms can be grouped based on the kinds of data available for the learning model: supervised learning, unsupervised learning, semi-supervised learning, and reinforcement learning.

\textit{Supervised Learning:}
The ML algorithms of this learning approach learn the relation between input and output via a mapping function using labeled data.
Each input sample in a training dataset is tagged with the answer (a.k.a., label), which allows the trained model to classify or predict the outcome for an unforeseen input sample~\cite{kotsiantis2007supervised}.
Supervised learning algorithms are usually used to handle classification problems (assign a sample in the test set into a discrete class) and regression problems (express the relationship between dependent and independent variables in continuous data).
Some regular supervised learning algorithms are decision tree, random forest, Naive Bayes, \textit{k}-nearest neighbor, and support vector machine (SVM).

\textit{Unsupervised Learning:}
Unsupervised learning involves the utilization of AI/ML algorithms for unlabeled data analysis and clustering.
These algorithms cannot be applied to classification and regression problems directly, but they are capable of modeling hidden patterns and finding out data groups without the need for human intervention.
Unsupervised learning algorithms (for example, hierarchical clustering, k-means clustering, principal component analysis, and association rule) can be used for some common tasks of data mining, such as clustering, association, and dimensionality reduction~\cite{hu2015unsupervised}.

\textit{Semi-supervised Learning:}
Semi-supervised learning is introduced to partly counter the disadvantages of supervised learning (e.g., expensive cost for labeling data by scientists and ML engineers) and unsupervised learning (e.g., the limitation of application spectrum). 
In semi-supervised learning, an AI model is trained upon the combination of labeled and unlabeled data.
A basic procedure of this type of learning involves two steps: clustering similar data using an unsupervised learning algorithm and then using the existing labeled data to label the remaining unlabeled data~\cite{sheikhpour2017survey}.
Some well-known algorithms for semi-supervised learning are graph-based model, generative model, boosting, and self-training~\cite{van2020survey}.

\textit{Reinforcement Learning}
Reinforcement learning (RL) is a group of ML algorithms for making a sequence of decisions, in which an agent learns to attain a goal in an uncertain and complex environment~\cite{kiumarsi2018optimal}. 
An AI machine should come through trial and error to reach a nearly optimal solution for a game-like scenario.
The object of an RL model is how to perform the task to maximize the reward and minimize the penalty, beginning with totally random trials and ending with sophisticated tactics and superhuman skills~\cite{chen2021deep}. 
By exploiting the power of the searching scheme with many trials, RL is one of the most effective ways to imply machine creativity.

\subsubsection{Advanced Techniques}
DL, a subset of AI and ML that develops multi-layered artificial neural networks to attain state-of-the-art accuracy in many classification and regression tasks, has been exploited for various applications in multiple domains~\cite{ho2021short, hua2020dran, pham2021intelligent}.
Unlike traditional ML techniques, DL can automatically learn underlying features of unstructured data without human intervention or human domain knowledge. 
The highly flexible architectures of DL allow learning systems to process raw data directly and improve learning performance when the data is provided enough.
Here we discover some well-known deep architectures, including recurrent neural network (RNN), convolutional neural network (CNN), self-organizing map (SOM), and autoencoders.

\textit{Recurrent neural network:}
RNN is one of the foundational neural network architectures from which various deep architectures, such as long short-term memory (LSTM) and gated recurrent unit (GRU) networks, are developed with some structural improvements.
Besides feed-forward connections in regular multilayer networks, an RNN has some feedback connections associated with the preceding layers.
The computing flow derived by the feedback connections allows RNNs to maintain memory of past inputs and process models in time~\cite{lalapura2022recurrent}. 
RNNs can be unfolded in time and trained with back-propagation mechanisms.

\textit{Convolutional neural network}
As one of the most successful deep network architectures, CNN leverages principles from linear algebra (especially matrix multiplication) to identify complex patterns from high-dimensional unstructured data~\cite{gonzalez2018deep}.
Early layers compute features from coarse to fine in a regular CNN, and later layers recombine these features into higher-level representations. 
CNNs are distinguished from other deep network architectures by their superior performance with different data types, including images, videos, audio signals, and communication signals.
There are three main layers in a CNN: convolutional layers for feature extraction, pooling layers for dimensionality reduction, and fully-connected layers for classification.
Several standard architectures were introduced to solve various challenging tasks in the computer vision domain: AlexNet, VGG, GoogleNet, ResNet, DenseNet, Inception, and EfficientNet.

\textit{Self-organizing map:}
SOM is an unsupervised neural network to find clusters of the input data points by reducing its dimensionality~\cite{ramos2020forbidden}.
In common SOM architectures, weights serve as a characteristic of the node.
At the beginning, the inputs are normalized and then randomly chosen to feed the network. 
Random weights close to zero are associated with each feature of the input record, which represents the input node.
The node with the least Euclidean distance (between each of the output nodes and the input node) is recognized as the most accurate input representation and denoted as the best matching unit (BMU). 
By establishing these BMUs as centroid, other units are calculated similarly and assigned to the shortest distance cluster.

\textit{Autoencoder:}
An autoencoder is a special type of neural networks, which is trained with the compression and decompression functions to map its input to output~\cite{jing2020self}.
In an autoencoder network, the input layer is encoded into the hidden layer using an encoding function for compression, where the number of hidden nodes is much less than the number of input nodes.
Accordingly, this hidden layer contains the compressed representation of the original input.
The output layer aims to reconstruct a decoding function for decompression input information.
The difference between the input and the reconstructed output in the training phase is calculated using an error function.
Since autoencoders can learn continuously with backward propagation, they are usually applied for self-supervised learning tasks~\cite{song2018self}.

\subsection{Role of AI in the Metaverse}

By merging AI with other technologies, such as AR/VR, blockchain, and networking, the metaverse can create secure, scalable, and realistic virtual worlds on a reliable and always-on platform.  
According to the seven-layer metaverse platform, it is undoubted to realize the important role of AI to guarantee the reliability of infrastructure and improve its performance so far. In the 5G and future 6G systems, many advanced ML algorithms with supervised learning and reinforcement learning have been adopted for different challenging tasks, such as efficient spectrum monitoring, automatic resource allocation, channel estimation, traffic off-loading, attack prevention, and network fault detection. 
With sensor-based wearable devices and other human-machine interaction gadgets, simple human movements and complex actions can be analyzed and recognized based on learning ML and DL models. 
Therefore, users' movements in the real world are projected into the virtual worlds, allowing users to fully control their avatars to interact with other objects in the metaverse comfortably.
Moreover, these avatars can engage with many modalities adopted in the real world, such as facial expressions, emotions, body movement, and physical interactions, besides speech recognition and sentiment analysis, which are powered by AI in terms of accuracy and processing speed.

\begin{figure*}[!t]
	\centering
	\includegraphics[width=170mm]{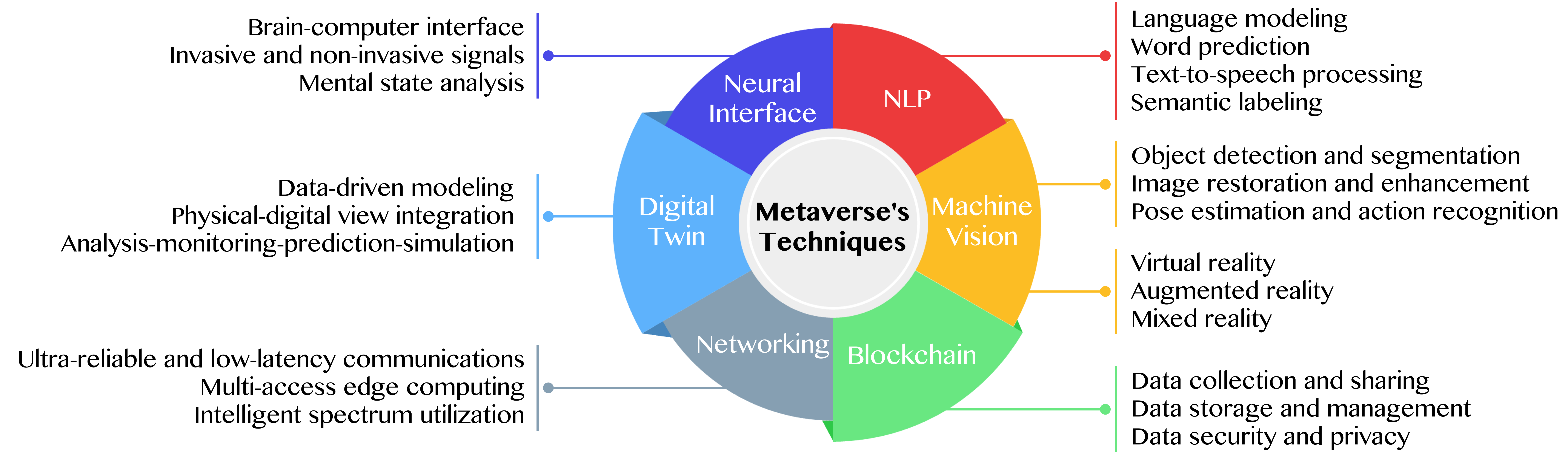}
	\caption{Primary technical aspects in the metaverse, in which AI with ML algorithms and DL architectures is advancing the user experience in the virtual world.}
	\label{fig_technical_metaverse}
\end{figure*}

Although XR/VR someway represents the facade of a metaverse with immersive devices like head-mounted displays, AI is a pivotal technology working behind the scenes to build a creative and beautiful world, thus bringing a seamless virtual-reality experience to users.
AI can facilitate the content creation process, for example, some AI modules like GANverse3D introduced by NVIDIA enable developers and creators to take photos of objects and then make virtual replicas. 
Several DL-based methods have been proposed for rendering 3D objects (including human body parts), which can achieve very impressive accuracy while presenting real-time processing accelerated by both software (e.g., PyTorch3D library from Facebook AI and TensorRT from NVIDIA) and hardware (e.g., GPUs) 
Meta just very recently introduced the AI research supercluster (RSC)~\cite{moore2022meta}, believed as among the world-class fastest AI supercomputer that will speed up AI research and be served for building the metaverse. Furthermore, RSC can help AI researchers and scientists develop better DL models from massive data, including text, speech, image, video, for various services/applications. Accordingly, any achievements and outcomes derived from RSC will be used as fabrics to build the metaverse platform, in which AI-driven products will be of considerable importance.

\section{AI for the Metaverse: Technical Aspect}
\label{sec_Technical}

This section investigates the state-of-the-art AI-based methods in six technical aspects: natural language processing, machine vision, blockchain, networking, DTs, and neural interface; which present the potential for the metaverse as shown in Fig.~\ref{fig_technical_metaverse}. Accordingly, the experience of users in the metaverse is enhanced significantly with nearly no boundary between the virtual world and the real world. 

\subsection{Natural Language Processing}
Natural language processing (NPL), also known as computational linguistics, encompasses a variety of computational models and learning processes to solve practical problems of automatically analyzing and understanding human languages, including speech and text. Besides, the field of NLS considers many topics, such as speech-to-text, text-to-speech, conversation design, voice branding, and multi-language and multi-cultural in voice. Furthermore, NLP plays a vital role in the metaverse regarding intelligent virtual assistants (a.k.a., chatbot). Particularly, NLP is principally responsible for enabling chatbots to understand complicated human conversation in the context of varying dialects and undertones. Empowered by AI, chatbots can answer nuanced questions and learn from interaction to improve the quality of responses. The AI chatbots are developed to assist users in some virtual environments like the metaverse.

As one of the most important tasks in NLP, language modeling predicts words or simple linguistic units by capturing syntactic and semantic relations of preceding words and units, which is useful for machine translation and text recommendation. 
In~\cite{daniluk2017frustratingly}, many neural networks with key-value attention mechanisms were built and evaluated on the Wikipedia corpus dataset to conclude that RNNs and LSTM networks with the attention mechanisms can outperforms large-scale networks while reducing memory in use.
In~\cite{benes2017residual}, a memory network with residual connection was designed to improve the performance of language modeling in terms of test perplexity if compared with regular LSTM~\cite{ jozefowicz2015empirical} having an equivalent size.
Some recent CNNs have been leveraged to address the long-term dependencies in long sentences and short paragraphs, especially being efficient to specific and complicated word patterns~\cite{pham2016convolutional}. Some deep networks were designed with advanced modules and connection structures to enhance language modeling efficiency, such as gated connection and bi-directional structure~\cite{liu2020chinese}.
Besides word-aware language models, many character-aware models have been introduced with AI algorithms to deal with various diversified languages in the world. Both the CNN and LSTM architectures~\cite{athiwaratkun2017malware} were applied to analyze the representation of words from characters as the input. Some models were evaluated on many datasets of English, German, Spanish, French, and Arabic, in which they showed the effectiveness in identifying prefixes and suffixes, recognizing hyphenated words, and detecting misspelled words~\cite{ma2020charbert, sharma2020deep, khysru2021tibetan}. Generally speaking, character-aware and word-aware modeling techniques allow natural language understanding systems to extract syntactic and semantic information for some common tasks in the metaverse, such as part-of-speech tagging, named-entity recognition, and semantic role labeling.

DL has been further exploited to overcome the learning limitation of conventional ML algorithms and effectively deal with many challenging tasks in NLP.
Some CNNs with sample and advanced architecture in~\cite{jin2020multi} were leveraged to cope with multiple sentence-based tasks, such as sentiment prediction and question type classification.
Moreover, sentiment analysis and recognition tasks may require feature extraction of aspects and sentiment polarities~\cite{wang2019tree}, which are potential to improve the reliability and flexibility of virtual assistant units in the metaverse.
Natural language generation is an advanced functionality of chatbot to generate reasonable task-specific conversation-oriented text.
Some single RNN/LSTM and mixture LSTM-CNN models were proposed to generate short text in image captioning and long text in virtual question answer~\cite{liu2019bfgan}.
Besides supervised learning, unsupervised and reinforcement learning with deep models have been adopted for some specific NLP tasks, such as text parsing, semantic labeling, context retrieval, language interpretation, and dialogue generation~\cite{young2018recent}.
In the metaverse, NLP techniques should be combined to fully provide text-based and speech-based interactive experiences between human users and virtual assistant.

\begin{figure*}[t]
	\centering
	\includegraphics[width=140mm]{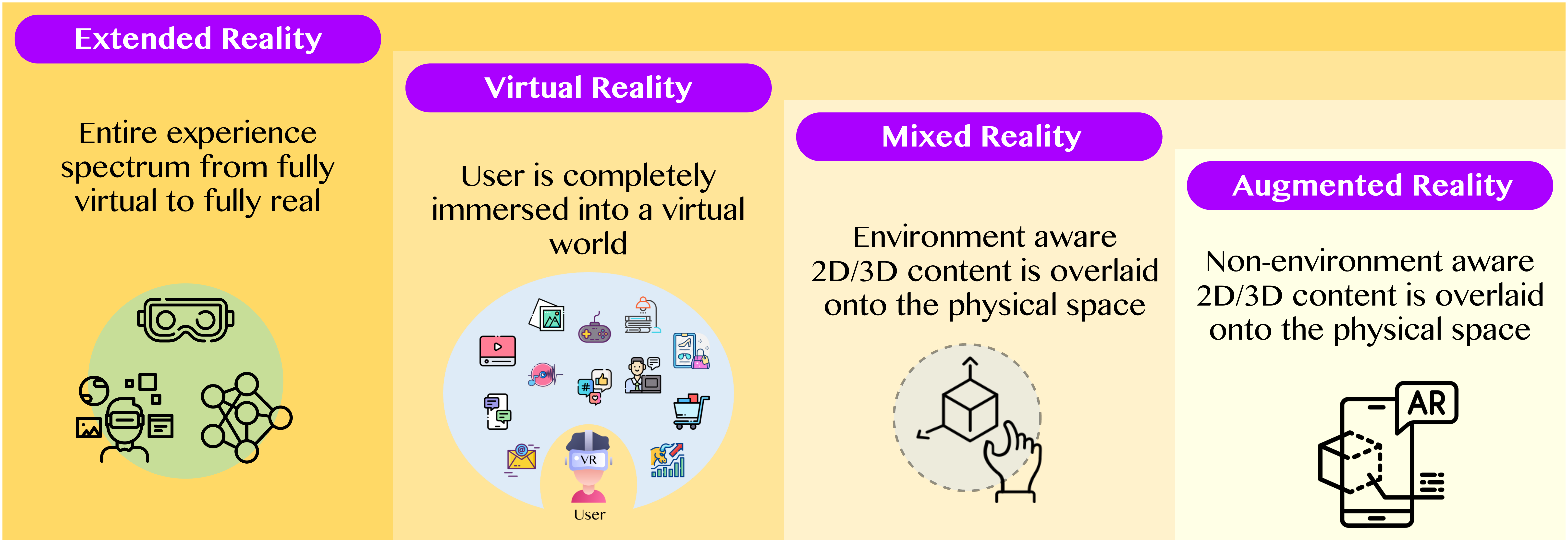}
	\caption{The difference between AR, MR, and VR under the umbrella of XR.}
	\label{fig_XR}
\end{figure*}

\subsection{Machine Vision}
Machine vision, including computer vision and XR in cooperation, is one of the central technologies to obtain the foundation of the metaverse. The raw data perceived from visual environments (via optical display and video player) is captured and processed to infer high-level information, which is then shown to users over head-mounted devices and others, such as smart glasses and smartphones. Indeed, computer vision allows XR devices to analyze and understand user activities based on visual-based meaningful information. Represented as avatars in the virtual worlds, the users can freely move in 3D maps and interact with virtual objects in the metaverse.

\subsubsection{Extended Reality}
XR is defined as an umbrella term that encapsulates VR, AR, mixed reality (MR), and everything in their gaps as shown in Fig.~\ref{fig_XR}. 
Although some revolutionary experiences are offered for VR and AR, the same original technologies are fueling the innovation and development of MR.  
While AR provides the experiences of graphics, video streams, and holograms in the physical world and VR offers viewing experiences in a fully immersive digital world, MR can deliver a transition experience between AR and VR.
Along with these reality technologies, human users can experience the metaverse and enjoy diversified services in both the physical and digital worlds. 
While XR and AI are distinct sectors, they can be combined to reach a fully immersive in the metaverse.

While conventional two-dimensional (2D) videos are limited by the small field of view (FoV), the 360-degree videos providing unlimited view-point with all directions are suitable for VR performance. 
Many commercial VR headsets are designed to satisfy the high-class using requirements, such as performance and comfortableness, which encompass multiple tasks driven by AI automatically. 
With the VR headset, a human user can experience various services and applications in the metaverse, and further create hyperreal media contents in the virtual world.
Some AI algorithms have been applied in VR devices to improve the human-machine interaction experience based on visual-based information.
For the prediction of user’s eye fixations in some gaze-based applications, such as content design and rendering, a DL framework with multiple CNNs~\cite{hu2021fixationnet} was built to deal with various kinds of input data, e.g., VR image, gaze data, and head data.
This model effectively exploited the correlations between eye fixations and other factors likes VR content and headset motion. 
In~\cite{pham2018human}, neural networks were adopted for human identification and authentication by analyzing periodic behaviors between users and VR gears (e.g., controllers and head-mounted display).
This work's effectiveness in delivering useful information and treatment recommendation was verified in collaboration and gaming scenarios and has shown some applicable potentials in other scenarios, such as working and shopping.
To enhance users' quality of experience (QoE) in the virtual world, an innovative human-machine interface approach~\cite{zhu2020sensory} was proposed by incorporating triboelectric sensory gloves and display components in VR devices to recognize multi-dimensional motion of gestures.
As a result, virtual objects which are recognized by leveraging ML/DL algorithms can be manipulated in the real-time VR/AR space.
To access the contents in the metaverse and interact with virtual objects in the digital world, not only AR headsets but also other devices (e.g., triboelectric gloves, hand-held touchscreen devices, and tabletops)~\cite{kataoka2019new} are taken into consideration regarding specific applications, services, and infrastructures.

To satisfy the service’s demands about high-resolution video viewing experiences with VR devices, it is necessary to develop an effective video quality assessment method, in which DL has represented as a powerful tool to obtain quantitative and qualitative benchmark objectives. In~\cite{wu2019virtual}, a high-performance method was developed for VR quality assessment by building a 3D CNN architecture, in which the video prediction results are validated via some common image quality assessment metrics without video reference. 
Compared with the baseline, which performed handcrafted feature extraction and ML algorithms, the 3D CNN-based approach showed the superiority in term of VR video quality assessment and benchmark~\cite{yang20183d}. 
Quality assessment was extended for 2D and 3D foveated-compressed videos in~\cite{jin2021subjective}, which allows VR systems to effectively handle the limited data transmission bandwidth. 
The advantages and limitations of current video quality assessment methods were analyzed and exposed in~\cite{han2020accuracy}, which can be useful to design an effective video transmission mechanism for various AR systems and diversified video contents. 
On the way to become the next mainstream for consumers and business, MR is defined as a blend of physical and digital worlds, which establishes the natural and intuitive interactions between 3D human, computer, and surrounding environment~\cite{barba2012here}. 
This new reality is activated by the recent revolutions in computer vision, graphic processing, display, remote sensing, and AI technologies~\cite{kersten2012dvv}.
Compared with VR and AR, MR has more potentials for the metaverse thanks to its hybrid physical-virtual experiences via two main types of devices\footnote{https://docs.microsoft.com/en-us/windows/mixed-reality/discover/mixed-reality}: holographic device with see-through display allows users to manipulate physical objects while wearing it and immersive device allows users to interact with virtual objects in the virtual world.
In the future, new devices to enhance the visual-interactive experiences of users in the metaverse should minimize the different gaps in terms of specification and utility between holographic devices and immersive devices.

\subsubsection{Computer vision}
In the last decades, computer vision has been empowered by AI, especially DL with a variety of network architectures to improve the overall accuracy of visual systems with efficient cost thanks to high-performance graphic processing units.
Some fundamental computer vision technologies are potential to enhance the experience of human users in the metaverse, thus enabling users in the physical world to interact with the virtual environment in the digital world smoothly.

Semantic segmentation and object detection are two fundamental tasks in the computer vision domain, where semantic segmentation categorizes each pixel in an image to be one of pre-defined semantic classes~\cite{liu2021visual} and object detection aims to localize all possible objects in an input image by drawing corresponding bounding boxes with object information in tag~\cite{huynh2016nic}. 
Early segmentation works mostly adopted local feature extraction and tracking in cooperation with classification algorithms, which were limited by unacceptable segmentation performance when dealing with large-messy datasets. Recently, numerous DL-based approaches have shown considerable improvement in terms of performance by exploiting different deep architectures if compared with traditional methods~\cite{minaee2021image}. The powerful capability of CNNs in extracting deep visual features at multi-scale image resolutions has been exploited to design advanced segmentation models in~\cite{long2015fully, shelhamer2017fully, chen2018deeplab, wei2016stc, zhou2019unet++}. For example, DeepLab~\cite{chen2018deeplab} leveraged atrous convolution to improve feature learning efficiency by enlarging the receptive field of filters while keeping a small number of parameters or a low computational cost. Because of learning classification models at pixel level, image segmentation usually consumes considerable computation and large memory~\cite{hua2020cross, hua2018convolutional}. To overcome this challenge, some efforts have been presented with some skills of network designs and learning techniques, such as transfer learning~\cite{liu2019structured}. The virtual environment in the metaverse is usually built with diversified visual units (e.g., single objects and multi-object modules); therefore, AI-based object detection must deal with a huge number of complicated classes, including real and virtual objects. Numerous recent object detection works have exploited CNN architectures to achieve impressive performance in terms of accuracy and processing speed~\cite{cheng2018improved, zhu2018attention, liu2020picanet}. DL-based semi-supervised and unsupervised learning models have been recommended to tackle unseen classes in the training dataset~\cite{tang2018visual}. Some natural problems of object detection in the 3D environment, such as occlusion, illumination variation, and view-point change, have been taken into consideration by incorporating advanced image processing and depth sensing algorithms~\cite{chen20183d, feng2021relation}. In this context, depth estimation can improve the accuracy of object positioning~\cite{lai2019real} in the 3D virtual world, but more geometric sensors are required to estimate depth information.

\begin{figure}[!t]
	\centering
	\includegraphics[width=88mm]{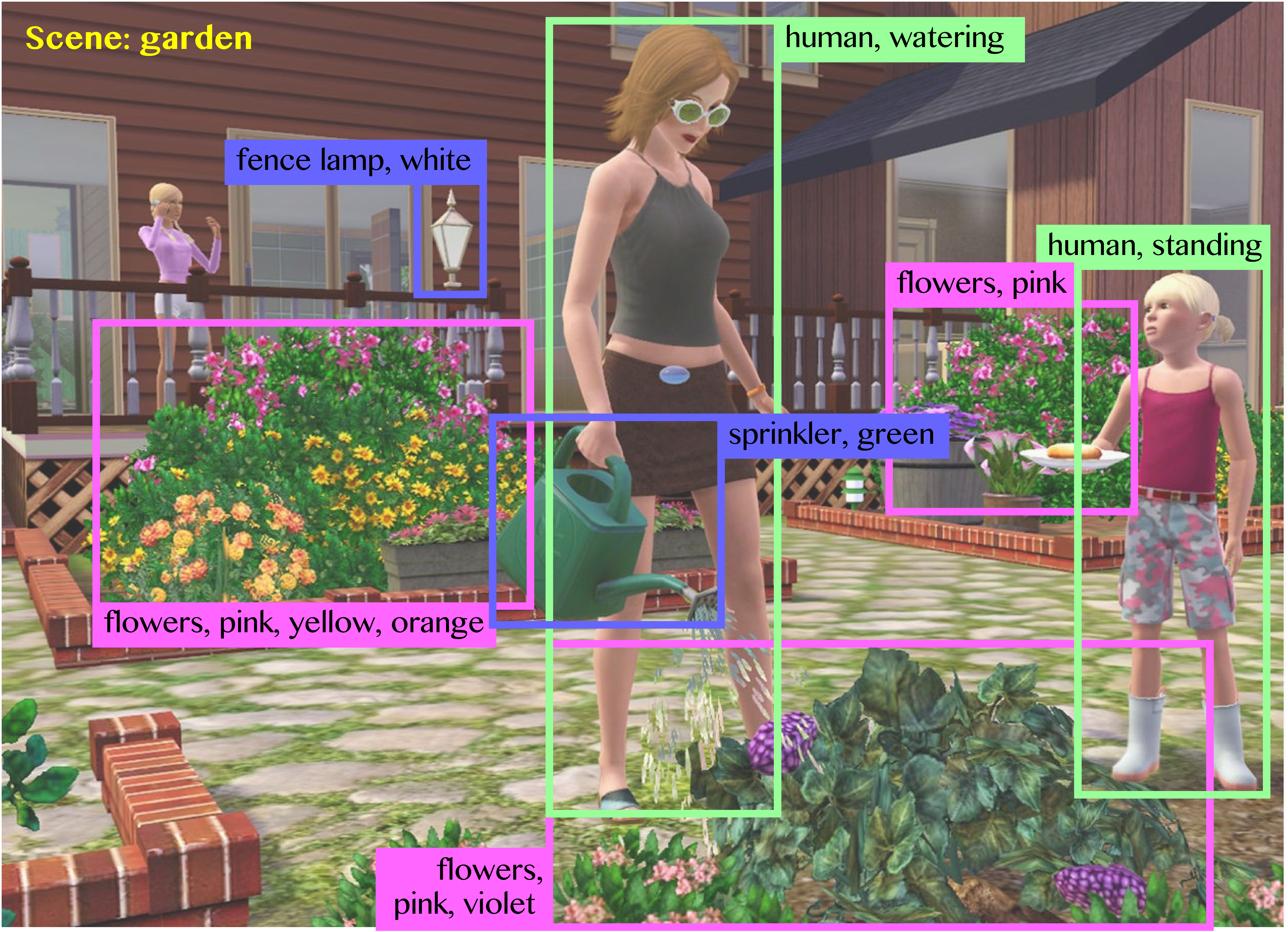}
	\caption{Computer vision in the metaverse with scene understanding, object detection, and human action/activity recognition.}
	\label{fig_computervision}
\end{figure}

In the virtual world, some image quality reduction problems, such as noise, blurring, and low resolution, should be addressed to enrich the visual perception of users. In the perspective of image processing and computer vision, these problems are studied over two tasks: image restoration and image enhancement.
In~\cite{yeh2019multi}, a decomposition-guided multi-scale CNN-based method was proposed to remove single image haze, in which deep residual structure and U-Net~\cite{ronneberger2015u} learning frame are combined to improve decomposed image components (so-called as feature maps) while avoiding color distortion. Some other advanced image restoration works exploited CNN architectures to reduce image compression artifacts, restore clean images from downscaled and blurred images, and reconstruct missing details~\cite{wang2018aipnet, jin2019flexible, zhang2020residual}. It is noted that the differences in terms of image quality and video specification between the clean virtual contents and real displayed images/videos can appear in VR devices~\cite{wang2019research, saalfeld2021vr}. These gaps can be filled effectively with AI-empowered image restoration methods, such as blur estimation, hazy removal, color correction, and texture reconstruction, but the computational complexity should satisfy the real-time video processing speed (usually measured by frames per second – FPS metric)~\cite{lahiri2020lightweight} to guarantee high-class user experience in the metaverse. Image enhancement has been widely considered for XR with some common tasks, such as contrast increment and super-resolution construction. In the past, many traditional image enhancement methods have been studied by applying image processing techniques, for example, histogram analysis and image decomposition~\cite{huynhthe2014using, zhang2019survey}. Recently, numerous impressive works of image enhancement have achieved considerable performance improvement by exploiting ML algorithms~\cite{syrris2015image}, especially DL with CNN architectures~\cite{wang2020improved, mei2020spatial, lee2020srnpu, yang2020hybrid}. For example, a convolutional down-sampling and up-sampling network was introduced in~\cite{wang2020improved} to improve the overall contrast of images, in which the deep features of RGB (red, green, and blue) channels are combined via a feature-based fusion scheme to obtain cross-channel contrast balance. In~\cite{dong2015image}, a fully CNN for image super-resolution was proposed with a lightweight structure, which can learn an end-to-end relation between input low-resolution images and output high-resolution images. Compared with some traditional sparse-coding-based methods~\cite{yang2010image}, this approach has shown the superiority in terms of image quality and processing speed. Super-resolution can become a cost-efficient solution that allows service providers to build high-resolution virtual worlds from low-resolution image/video sources. 

In the metaverse, play users can control their avatars (or virtual characters) and interact with other users or non-player characters (NPCs), in which the posture and action of avatars should be estimated and recognized automatically with the support of motion sensing interactive devices, such as controllers, gloves, and cameras~\cite{dang2019deep}.
While human pose estimation aims to identify the body parts (or key body joints of a skeleton) and then track them in the real-time environment~\cite{zheng2020deep}, action recognition allows systems to understand single actions and complex interactive activities (e.g., human-machine interaction and human-human interaction)~\cite{ramanathan2014human}. 
To deal with the problem of estimating human pose in cluttered environments, two discriminative models based on standard structural support vector regression (SVR) and latent structural SVR were studied in~\cite{chen2011human}, which are capable of extracting structural dependencies as the correlations between local features regarding pose representation.  
To improve the accuracy of body part localization and deal with varying views, the depth information acquired by depth cameras has been learned along with color information by advanced ML and DL models~\cite{rogez2019lcr, he2019interacting, zhao2020learning}.
In these works, some CNN architectures with high-class structural connections, such as dense layer connection, skip connection, and channel-attention connection, were designed to estimate skeletal joints precisely besides addressing some challenging problems in computer vision like object occlusion.
In the line, human pose estimation has a close relation to action recognition, where the captured body information is useful to identify actions via pattern recognition models.
Instead of detecting instant posture that can expose high confusion, many current works have tracked body motion in the temporal domain for a long-term observation to improve the accuracy of action recognition.
For example, some generative statistic models have been developed in~\cite{huynh2016interactive, tu2018ml, huynh2018hierarchical} to analyze human pose transition by capturing the spatio-temporal geometric features between different body parts.
Notably, the last decade has witnessed the revolution of visual-based action recognition with DL to significantly improve recognition accuracy and effectively deal with numerous realistic single actions and grouped activities~\cite{wang2017beyond}. 
Some methods proposed innovative networks with advanced CNN architectures~\cite{huynh2020encoding} and hybrid CNN-RNN architectures~\cite{wang2017beyond} to improve learning efficiency of action discrimination models.
Additionally, hand gesture recognition, gait identification, and eye tracking~\cite{huynh2020learning, hu2020dgaze} have been considered to improve interactive experiences in XR environments. 

\subsection{Blockchain}
In general, blockchain is defined as a digital ledger that contains a list of recorded transactions and tracked assets interconnected in a business network by using cryptography techniques. 
Blockchain can provide immediate, shared, and transparent information stored in an immutable and impenetrable ledger which can be accessed by only the network members with permission~\cite{gadekallu2022blockchain}. 
A typical blockchain network can track orders, payments, accounts, and other transactions. 
In the metaverse, a large amount of data (e.g., videos and other digital contents) is acquired by VR devices, transmitted over networks, and stored in data center without any security and privacy protection mechanisms, which can become the sensitive target of cyberattacks. In this context, blockchain with several unique features reveals a promising solution for security and privacy issues in the metaverse~\cite{cannavo2021howblockchain}, especially when it is empowered by AI technologies.
Besides, many creative activities and events offered by service providers to users will yield numerous in-metaverse objects/items (a.k.a., digital assets) which should be recorded and tracked via transparent transactions with smart contracts in blockchain.

\begin{figure}[!t]
	\centering
	\includegraphics[width=88mm]{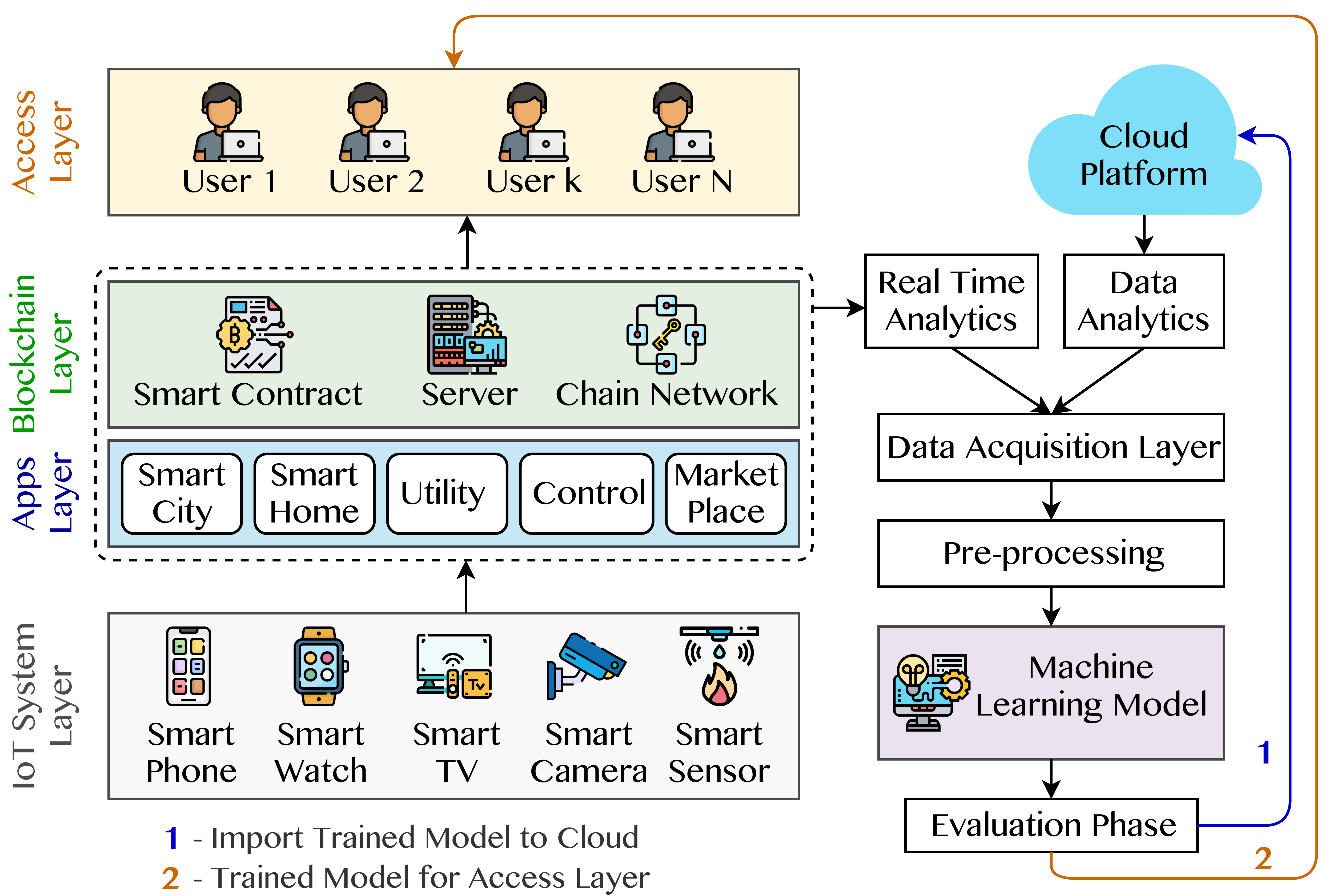}
	\caption{A blockchain-based IoT framework with ML to enhance security and privacy.}
	\label{fig_blockchainAI}
\end{figure}

In the last decade, numerous advanced methods for data acquisition, storage, and sharing have been proposed by combining blockchain and AI technologies in various application domains to obtain high data security and privacy~\cite{liu2020blockchain}, which have shown great potential to be deployed in the metaverse.
In~\cite{tanwar2019machine}, various conventional ML algorithms (e.g., clustering, SVM, and bagging) and innovative DL architectures (e.g., CNN and LSTM) were investigated for data analytics to detect and classify cyberattacks in blockchain-based networks. Some other concerns were also considered in this work, such as incentive mechanisms to encourage users to contribute authenticated data, AI-based smart contract evaluation, and cost-efficient model learning in on-chain environment. 
For the Internet of Things (IoT)-aided smart cities, an effective privacy-preserving and secure framework~\cite{kumar2021ppsf} was introduced by integrating blockchain with enhanced proof of work and ML with data transformation, which in turn robustly deals with various cyberattacks in smart city networks.
In~\cite{khan2021amachine}, deep extreme learning machine was exploited in a resource-efficient blockchain-based IoT framework, which improved the system security and privacy based on data interpretation and abnormality prediction.
This framework (see Fig.~\ref{fig_blockchainAI}) has shown high performance of fraud detection and threat prediction, and can be extended for dealing with security and privacy problems in data storage and sharing instead of data collection. 
Recently, DL has replaced traditional ML in terms of cooperating with blockchain to solve some challenging security and privacy issues of big data, where five essential characteristics of big data (i.e., velocity, volume, value, variety, and veracity) are presented.
For example, DeepChain~\cite{weng2021deepchain}, a CNN-based blockchain framework, was developed to ensure the privacy and integrity of data contributed by participants in a network.
Deep RL was exploited to achieve a secure mobile offloading in multi-access edge computing (MEC) based blockchain networks~\cite{nguyen2020privacy} and to obtain a secure vehicular crowdsensing in the blockchain-based Internet of vehicles (IoV) systems~\cite{wang2021secure}.

Federated learning (FL) has recently emerged as an effective solution to address the privacy problems of data sharing, in which multiple users train AI models with their own local data and collaboratively learn a global model at the server by a parameter aggregation mechanism.
In~\cite{lu2020blockchain}, FL was applied to address privacy issues in data sharing among multiple untrusted parties in a blockchain network.
This work integrates FL into a proof of training quality, a novel consensus mechanism, of permissioned blockchain to reduce computing and communication costs.
To guarantee high privacy of massive data generated by heterogeneous IoT devices, FL was deployed in a blockchain-based resource trading system~\cite{fan2021hybrid}. 
A smart contract-based incentive algorithm was proposed to encourage edge nodes to contribute and evaluate FL tasks.
For vehicular edge computing in intelligent transportation systems, FL was combined with blockchain to collaboratively detect malicious attacks~\cite{liu2021blockchain}.
While FL can offload the trained intrusion detection model to distributed edge devices to reduce computing resources of the central server, blockchain can ensure the security of the aggregation model in both the model storage and sharing processes.
Besides data security and privacy, interoperability is another important concern in blockchain to collaborate with different parties using different data infrastructures.
For example, a learning analytics framework~\cite{ocheja2018connecting} was studied to obtain solid interoperability between multiple blockchain participants which have to share a single ledger.
Integrating blockchain into FL were recently found in computing resource allocation and management applications~\cite{umer2021stbfl,li2021blockchain} to address various problematic issues in centralized systems, such as external cyberattacks, server malfunctions, and untrustworthy server.
In the metaverse, wherein multiple parties join and contribute digital content having different formats and structures, data security, privacy, and interoperability can be fully handled by collaboratively developing blockchain and AI.

\subsection{Networking}

The metaverse serves a massive number of users regarding pervasive network access over wireless networks.
In the last decade, several innovative technologies have been introduced to improve the overall performance of wireless communication and networking systems, in which AI has been intensively used at multiple layers of a network architecture~\cite{chen2019artificial}.
Real-time multimedia services and applications in the metaverse usually demand a reliable connection with high throughput and low latency to guarantee a basic-level user experience at least.
As the requirements of fifth-generation (5G) networks, the peak data rate should be around 10 Gbps (gigabits per second) and the end-to-end delay cannot exceed 10 ms (millisecond). In this context, ultra-reliable and low-latency communications (uRLLC) represent the foundation to enable the development of emerging mission-critical applications~\cite{she2021atutorial}. Several optimization algorithms have been introduced to achieve uRLLC in 5G networks and beyond (e.g., sixth-generation 6G), but most of them require high computing resources. ML and DL have shown great potential to effectively handle existing challenging tasks, such as intelligent radio resource allocation~\cite{she2020deeplearning}, in 5G/6G networks while meeting a very low latency. 
RL was leveraged to address the resource slicing problem for enhanced mobile broadband (eMBB) and uRLLC~\cite{alsenwi2021intelligent}, in which the complicated patterns of resource allocation and scheduling are formulated to collaboratively learn network states and channel conditions.
In another work~\cite{gu2021deepmultiagent}, RL showed the effectiveness in terms of joint subcarrier-power management and allocation, thus significantly reducing latency and improving reliability on the Internet of controllable things.
Particularly, this work proposed a double Q-learning network to optimize the total spectrum efficiency via subcarrier assignment and power control policy, and accelerate learning convergence.
As a vital role to enable uRLLC, efficient radio resource management was investigated in~\cite{azari2019riskaware} with a distributed risk-aware ML approach to monitor and manage the transmission of non-scheduled and scheduled uRLLC traffics.

Lately, DL has been exploited for many tasks in uRLLC, including spectrum management, channel prediction, traffic estimation, and mobility prediction. 
Two advanced CNN architectures, namely MCNet~\cite{huynh2020mcnet} and SCGNet~\cite{tunze2020sparsely}, were designed in physical layer to automatically identify the modulation types of incoming signals, which in turn allows the receiver to demodulate accurately and enhance the spectrum utilization efficiency accordingly.
To overcome the high computational cost of conventional channel state information (CSI) estimation approach, an online CSI prediction method~\cite{luo2020channel} was proposed a supervised learning framework by combining CNN and LSTM, in which two-stage training mechanism was deployed to improve the robustness and stableness of CSI estimation in practical 5G wireless systems.
In~\cite{guo2019deepspatial}, an end-to-end CNN architecture was designed with 3D convolution for intelligent cellular traffic forecasting, in which the deep model can learn the underlying correlations of traffic data in both the short-term and long-term spatial patterns.
Besides achieving high accuracy of traffic prediction, the deep network showed the effectiveness with different real-world scenarios, such as traffic congestion data and crowd flow data.
In conclusion, with ML algorithms and DL architectures, AI is a powerful tool to address many challenging problems of uRLLC in future wireless networks, which allows users to experience high-class integrated services in the metaverse with the guarantee of high throughput low latency. 
\begin{figure}[!t]
	\centering
	\includegraphics[width=88mm]{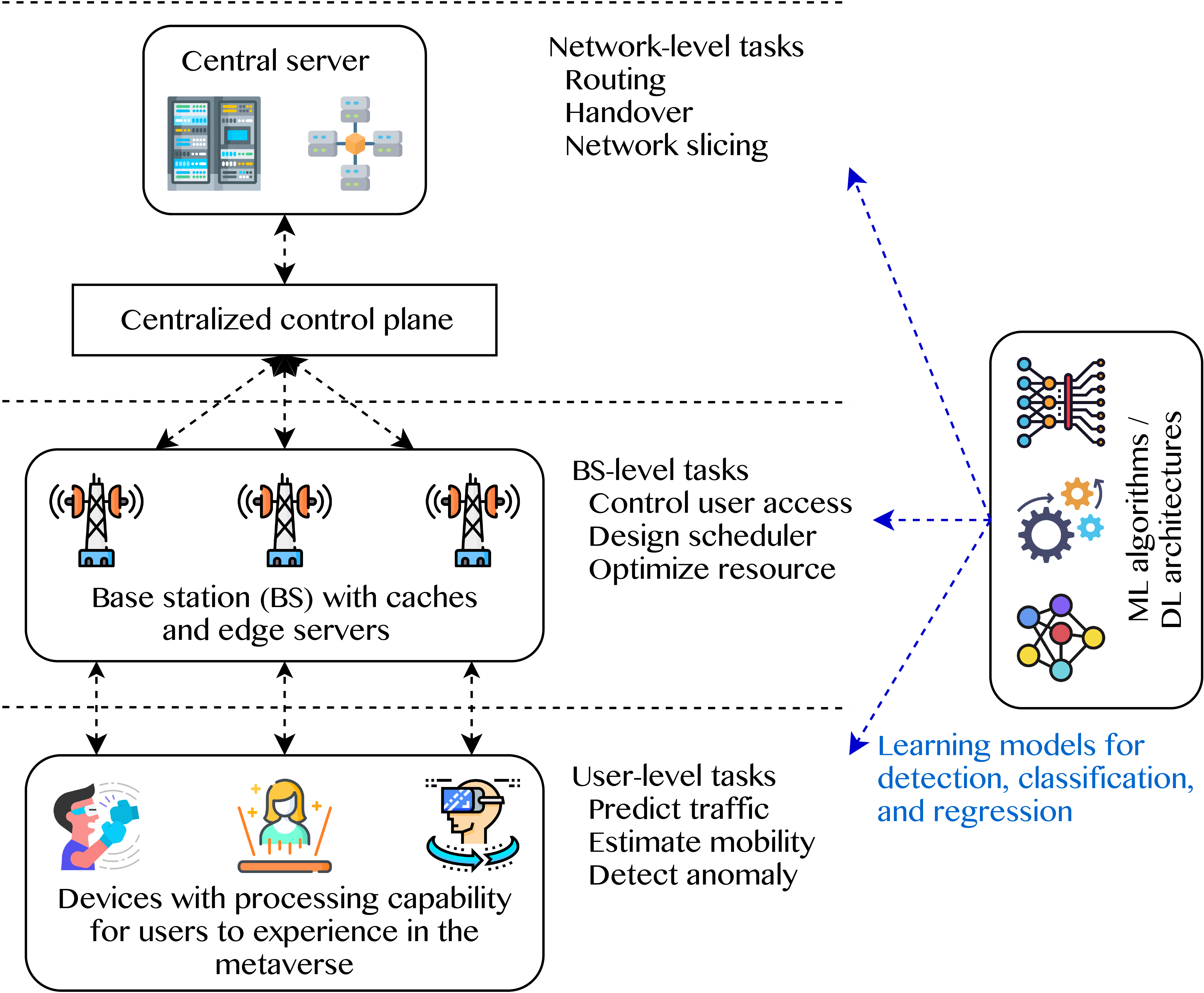}
	\caption{A general architecture in 5G and beyond for metaverse services and applications, in which AI with ML algorithms and DL models contribute in multi-level tasks.}
	\label{fig_networking}
\end{figure}

\subsection{Digital Twins}
As a digital representation of real-world entities, a DT can synchronize operational assets, processes, and systems with the real world along with some other regular actions, such as monitoring, visualizing, analyzing, and predicting~\cite{tao2019digital}. DTs are at the central of where the physical world and the virtual world interact via IoT connections~\cite{chen2021digital}; and therefore, any change in the real world will be rejected in the digital representation. With these distinctive properties, DT is found as one the fundamental building sectors of the metaverse and plays as the gateway for users to enter and enjoy services in the virtual world by creating exact replications of reality, including structure and functionality. For example, technicians can maneuver 3D representations of complex systems at multi-level sophistications (i.e., descriptive, informative, predictive, comprehensive, and autonomous) for a wide spectrum of purposes, such as technical training and commercial customization. Accordingly, DTs allow application developers and service providers to reconstruct virtual replications of machines and processes, in which any kind of physical analysis can be done remotely with AI~\cite{rathore2021therole}.

\begin{figure}[!t]
	\centering
	\includegraphics[width=88mm]{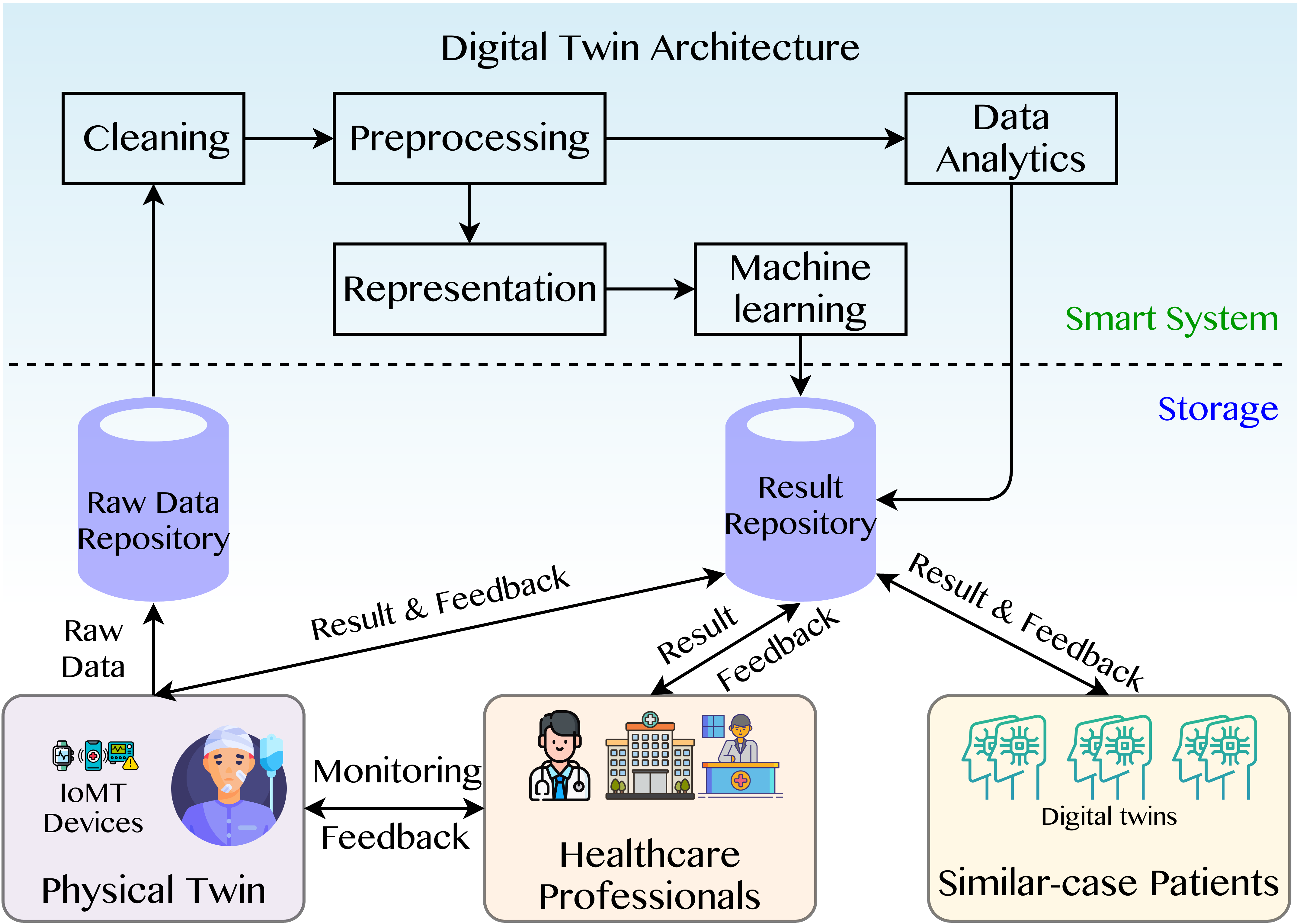}
	\caption{A data-driven DT framework for intelligent healthcare systems using ML to process raw data of IoMT devices.}
	\label{fig_digitaltwin}
\end{figure}

For industry 4.0, a reliable DT framework~\cite{darvishi2021sensor} was proposed for sensor-fault detection, isolation, and accommodation, in which a multi-purpose ML method with multi-layer perception neural network was deployed to validate sensory data, estimate fault condition, and identify fault sensor.  
As a digital replication to operate human-robot welding actions, a DT system in~\cite{wang2021digital} was developed along with VR and AI technologies to monitor and analyze welder behaviors. A generic ML framework with domain transformation, feature engineering, and classification was applied to recognize proper welding behaviors based on the data acquired from a bi-directional flow between robot and VR.
In~\cite{elayan2021digital}, a data-driven-based DT framework (see Fig.~\ref{fig_digitaltwin}) was studied to improve health diagnosis performance and promote better health operation in intelligent healthcare systems. DTs contributed in different phases to create virtual replications of patent health profiles, carry out collaborative activities of health professionals, and formulate a universal treatment plan for patients in same cases. ML models were built to learn meaningful information from raw data collected by the Internet of Medical Things (IoMT) devices to early detect health abnormalities and precisely recognize health problems. In another work that proposed a cyber-physical framework for smart urban agriculture services and applications~\cite{ghandar2021adecision}, DTs were designed to replicate a virtual representation of farming production, in which the sensory data acquired by practical sensors was processed by ML algorithms in a decision support system. To adapt to different types of product magnification, the DTs were built from small functioning modules to a whole process twin.

With a great capability of automatically learning features from high-dimensional unstructured data and effectively dealing with spatiotemporal learning models, DL has been recently applied in DT architectures for different services and applications. In~\cite{xu2022service}, a DT architecture was developed for the edge computing-aided Internet of vehicles (IoV) to improve the utilization efficiency of vehicle’s computational resources. To overcome the overload problem of edge devices, deep Q-network optimized the function approximation of DL and RL. For the performance investigation of uRLLC services and delay tolerant services in mobile edge computing systems, a DT framework was built in~\cite{dong2019deep} by replicating a virtual pattern of the real network environment. Remarkably, DL with feedforward neural network architecture was carried out to deal with varying network parameters of real-world phenomena. For industrial IoT, the work in~\cite{sun2021adaptive} built DTs to simulate and capture the operation state and real-time behavior of industrial devices, which were map into a digital world. To address the bias between real entity and its digital replication, a trusted-based aggregation with FL was carried out with a deep RL model to general improve performance while meeting resource constraints. 
With AI as a powerful analytics tool, DT can improve system performance, reduce process-related incidents, minimize maintenance costs, and optimize business and production. In addition, DT allows users to view the metaverse as an advancing replication of reality with full real-time synchronization from the physical world.

\subsection{Neural Interface}

\begin{figure}[!t]
	\centering
	\includegraphics[width=88mm]{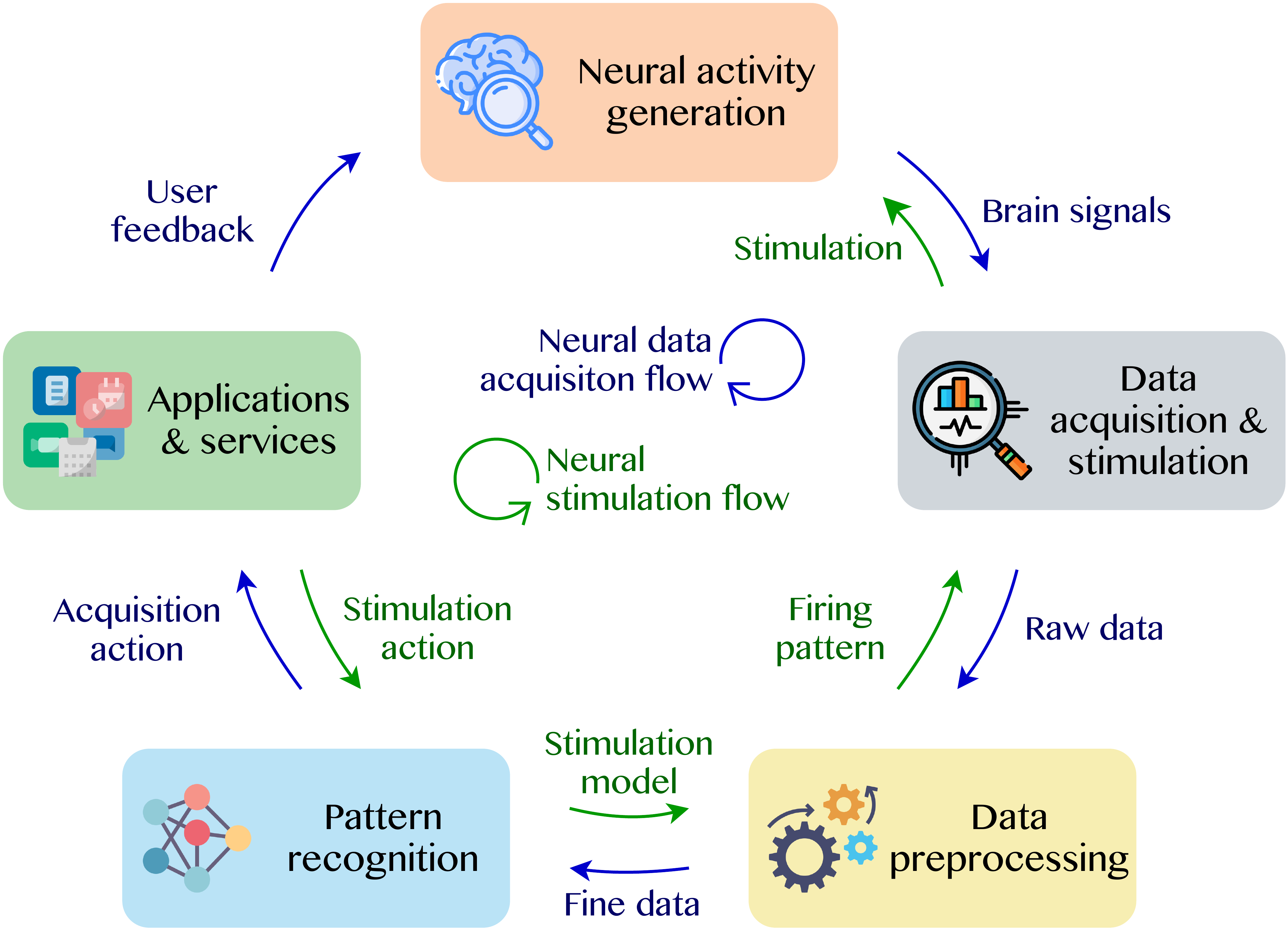}
	\caption{A common BMI cycle with primary components for processing neural signals and responding neural stimulations.}
	\label{fig_neuralinterface}
\end{figure}

Technology is definitely enriching the world around us by enhancing the human experience and fully filling gap between reality and virtual world in the metaverse. In this context, the most immersive popular interface to interactive with the virtual work is a VR headset with a controller. Many technology companies currently pay attention to neural interfaces, so-called brain-machine interfaces (BMIs) or brain-computer interfaces (BCI), that go beyond VR devices. The BMIs help to nearly clean the borderline between human and wearable devices. Many BMIs detect neural signals using external electrodes or optical sensors that adhere to the skull and other parts of the human body. According to these noninvasive devices, which only read and control mind at a rudimentary level, BMIs can manipulate thoughts with transcranial electromagnetic pulses. Fig.~\ref{fig_neuralinterface} describes a common BMI cycle with primary components for processing neural signals and responding neural stimulations~\cite{bernal2022security}. Besides data engineering techniques in the preprocessing stage, AI/ML algorithms in the pattern recognition stage enable analysis of complicated and sensitive neural signals accurately.

With electroencephalogram (EEG) signal as one of the most popular inputs of BCI systems, the work in~\cite{he2020transfer} studied a brain signal classification by two learning approaches: one is offline unsupervised classification and another is simulated online supervised classification. Besides that, two approaches achieved lower computational cost and better performance in common tasks, e.g., motor imagery, mental analysis, and event-related potential, the offline unsupervised mechanism did not require signal labeling for new subjects in the learning phase.
As building an accurate predictive model in BCIs to decipher brain activities into communication and control commands, the work in~\cite{abibullaev2019learning} learned discriminative spatiotemporal features to seize the most relevant correlations between different neural activities from EEG signals. Based on the reconstructed signal waveforms containing dominant frequency characteristics as feature vectors with lower dimensionality, several ML algorithms, e.g., logistic regression (LR), Naive Bayes, and SVM, were applied to investigate the performance of ERP.
In~\cite{matran2016brain}, the feasibility of using visual hemisphere to extract relevant information about the spatial location of targets in aerial images was investigated with feature selection and SVM classification, which were deployed in a rapid serial visual presentation (RSVP) procedure. Concretely, by learning ERP patterns from extracted discriminative features of EEG signals, the concerning target and its location in aerial images can be identified.
As the effort to increase the correct classification rate of EEG signals in BCIs, an advanced ML framework was introduced in~\cite{lv2021advanced} by combining an improved common spatial pattern algorithm and a transfer learning mechanism. Besides achieving high accuracy of classifying left-hand and right-hand imaginary movements, the trained AI model can be used for other classification and recognition tasks in the same domain via knowledge transferring technique.

Based on the superiority of capsule network (CapsNet) compared with traditional neural networks in terms of feature extraction and feature explanation, the work in~\cite{ma2021capsule} applied CapsNet to improve the accuracy of ERP detection in BCI systems. With highly discriminative spatial features and key temporal correlations extracted from EEG signals by capsule layers, CapsNet not only outperformed some state-of-the-art learning models (e.g., linear discriminant analysis and CNN~\cite{sakhavi2018learning}) but also obtained the practicality with different common spellers in the cognitive neuronscience domain.
In~\cite{jeong2020brain}, a hybrid DL framework was proposed with multi-directional CNN and bidirectional LSTM in an accurate brain-controlled robotic arm system. This proposed learning approach effectively calculated underlying spatial signal correlations in time, boosting the decoding performance for 3D multi-directional arm-based object grasping tasks.
Inspired by GoogleNet~\cite{szegedy2015going}, the work in~\cite{santamaria2020eeg} proposed EEG-Inception, a novel CNN for EEG-based classification tasks in BCI systems, which involved multiple inception modules in improving feature learning efficiency. Furthermore, an effective training strategy that incorporated cross-subject transfer learning and fine-tuning to reduce calibration time of ERPs and demonstrate the feasibility in real-world assistive applications.
In the future, brain-computer interfaces will truly promote the ultimate immersive interaction between the reality and the virtual world in the metaverse via consumer-ready mind-control systems.
The existing AI-enabled works on the subject of six concerning technical aspects, which are promisingly for the metaverse, are summarized in Table~\ref{tab_technicalAspect}.

\begin{table*}[ht!]
\setlength{\tabcolsep}{4pt}
\centering
\caption{Summary of AI For The Metaverse in The Technical Aspect.}

\begin{tabular}{|p{2.10cm}|p{0.75cm}|p{5.75cm}|p{7.85cm}|}
\hline \hline
\textbf{Technical Aspect} & \textbf{Ref} & \textbf{Task} & \textbf{AI Technique} \\ \hline \hline
NLP 
& \cite{daniluk2017frustratingly} & \multirow{3}{5.75cm}{Word and linguistic prediction for language modeling.} & 	RNNs and LSTM networks with the attention mechanisms.\\ \cline{2-2} \cline{4-4}
& \cite{benes2017residual}		& & Advanced memory network with residual connection. \\ \cline{2-2} \cline{4-4}
& \cite{liu2020chinese}		& & Deep networks with gated connection and bi-directional structure.  \\ \cline{2-4}
& \cite{athiwaratkun2017malware}	& Analyzing and understand the representation of words from characters	& General deep networks with CNN and LSTM architectures.\\ \cline{2-4}
& \cite{sharma2020deep}	& Identifying prefixes and suffixes and detecting misspelled words & DL framework with CNN, Bi-LSTM, and conditional random field. \\ \cline{2-4}
& \cite{jin2020multi} & Sentiment prediction and question type classification. & Various CNNs and LSTM networks with simple structures and advanced-designed architectures. \\  \cline{2-4}
& \cite{liu2019bfgan} & Generate short text in image captioning and long text in virtual question answer. & DL framework with single RNN/LSTM and mixture LSTM-CNN models. \\  \cline{2-4}
& \cite{young2018recent} & Semantic labeling, context retrieval, and language interpretation. & Unsupervised and reinforcement learning with common RNN/LSTM and CNN models. 
\\ \hline\hline

Vision Machine
& \cite{hu2021fixationnet} & Forecasting eye fixations in VR task-oriented virtual environment. & A DL framework with multiple CNNs for feature extraction and prediction. \\ \cline{2-4}
& \cite{wu2019virtual}	& \multirow{2}{5.75cm}{VR quality assessment for 2D and 3D foveated-compressed videos} &	\multirow{2}{7.75cm}{DL framework with CNNs architecture, in which 3D convolutional layers are built in feature extraction blocks.} \\
& \cite{jin2021subjective} & &  \\ \cline{2-4}
& \cite{hu2021fixationnet} & Forecasting eye fixations in VR task-oriented virtual environment. & DL framework with multiple CNNs for feature extraction and prediction. \\ \cline{2-4}

& \cite{chen2018deeplab}	& \multirow{2}{5.75cm}{Semantic segmentation and object detection.} & 	CNNs with atrous convolution. \\ \cline{2-2} \cline{4-4}
& \cite{hua2020cross}		& & CNNs with channel-wisely and spatially attentional schemes. \\ \cline{2-2} \cline{4-4}
& \cite{liu2020picanet}		& & CNNs with pixel-wise local and global attention pooling-convolution.\\\cline{2-2} \cline{4-4}
& \cite{tang2018visual}		& & CNNs with knowledge transfer via visual similarity and semantic relatedness.\\\cline{2-2} \cline{4-4}
& \cite{feng2021relation}		& & CNNs with 3D object-object relation graphs. \\ \cline{2-4}

& \cite{yeh2019multi}	& \multirow{2}{5.75cm}{Image/video quality enhancement (e.g., hazy removal, color correction, texture reconstruction, and super-resolution)} & 	Decomposition-guided multi-scale CNN architectures with deep residual structure and U-Net learning frame.\\ \cline{2-2} \cline{4-4}
& \cite{wang2020improved}		& & Combination of deep features extracted by CNNs via a feature-based fusion scheme.\\ \cline{2-2} \cline{4-4}
& \cite{mei2020spatial}		& & Full 3D CNN architectures with simultaneous and separated spatial-spectral joint feature learning mechanisms.\\ \cline{2-4}

& \cite{chen2011human}	& \multirow{2}{5.75cm}{Human pose estiation and action/activity recognition.}	& Discriminative model with latent structural SVRs. \\ \cline{2-2} \cline{4-4}
& \cite{rogez2019lcr}		& & CNNs with dense layer connection and channel-attention connection.\\ \cline{2-2} \cline{4-4}
& \cite{huynh2018hierarchical}		& & Generative models with latent Dirichlet and Pachinko allocations.\\ \cline{2-2} \cline{4-4}
& \cite{huynh2020encoding}		& & Advanced CNNs with geometric feature transformation. 
\\  \hline\hline
  
Blockchain 

& \cite{tanwar2019machine}	& Detection and classification of cyberattacks in blockchain-based networks.	& Conventional ML algorithms (e.g., clustering, SVM, and bagging) and DL architectures (e.g., CNN and LSTM). \\  \cline{2-4}
& \cite{khan2021amachine}	& Resource management in blockchain-based IoT framework.	& Deep extreme learning machine.\\  \cline{2-4}
& \cite{fan2021hybrid}	& Preservation of data privacy of heterogeneous IoT devices.	& FL framework with deep model using average aggregation mechanism. \\  \cline{2-4}
& \cite{liu2021blockchain}	& Detection of malicious attacks in intelligent transportation systems.	& FL framework with CNN model averaging and training.
 \\  \hline\hline
  
Networking 
& \cite{she2020deeplearning}	& Resource slicing problem for eMBB and uRLLC.	& RL with a policy gradient based actor-critic learning mechanism.\\  \cline{2-4}
& \cite{gu2021deepmultiagent}	& Subcarrier-power management and allocation.	& RL with double Q-learning network. \\  \cline{2-4}
& \cite{azari2019riskaware}	& Management of the transmission of non-scheduled and schedule uRLLC traffics.	& Supervised learning framework with conventional ML algorithms.\\  \cline{2-4}
& \cite{huynh2020mcnet}	& Automatic modulation classification in wireless systems.	& Advanced CNN architecture with sophisticated-designed modules of convolutional layers.\\  \cline{2-4}
& \cite{luo2020channel}	& Prediction of CSI in 5G wireless systems.	& A supervised learning framework by combining CNN and LSTM with two-stage training mechanism\\  \cline{2-4}
& \cite{guo2019deepspatial}	& Forecasting intelligent cellular traffic.	& An end-to-end CNN architecture was designed with 3D convolution.
 \\  \hline\hline
  
Digital Twin 

&\cite{darvishi2021sensor}	&Sensor-fault detection, isolation, and accommodation	& A multi-purpose ML method with multi-layer perception neural network.\\  \cline{2-4}
&\cite{elayan2021digital}	&Diagnosis of heart disease and detection of heart problems.	& A data-driven-based analysis framework with traditional classification algorithms.\\ \cline{2-4}
&\cite{ghandar2021adecision}	&Prediction of complete future system states relying DT of urban farming.	& Several common ML algorithms for regression (LR) and classification (SVM).\\ \cline{2-4}
&\cite{xu2022service}	&Improvement of resource utilization in edge computing-aided IoVs network.	& Deep Q-network optimized the function approximation of DL and RL.\\ \cline{2-4}
&\cite{sun2021adaptive}	&Simulation of the operation state and analysis real-time behavior via DT.	& FL framework with deep RL model. \\  \hline\hline
  
Neural Interface 

& \cite{abibullaev2019learning} & ERP classification in BCI systems. & Spatiotemporal feature extraction and ML algorithms (e.g., LR, Naive Bayes, and SVM). \\ \cline{2-4}
& \cite{matran2016brain} &  Spatial object localization in aerial images. & Feature selection with SVM classification. \\ \cline{2-4}
& \cite{ma2021capsule} & ERP detection in BCI systems. & Capsule network with primary capsule components. \\ \cline{2-4}
& \cite{jeong2020brain} & Brain-controlled robotic arm system. & DL framework with multi-directional CNN and bidirectional LSTM. \\ \cline{2-4}
& \cite{santamaria2020eeg} & EEG-based classification tasks in BCI systems. &  EEG-Inception network with cross-subject transfer learning and fine-tuning.   \\  \hline\hline
\end{tabular}%
\label{tab_technicalAspect}
\end{table*}

\section{AI for the Metaverse: Application Aspect}
\label{sec_Application}

\begin{figure*}[!t]
	\centering
	\includegraphics[width=180mm]{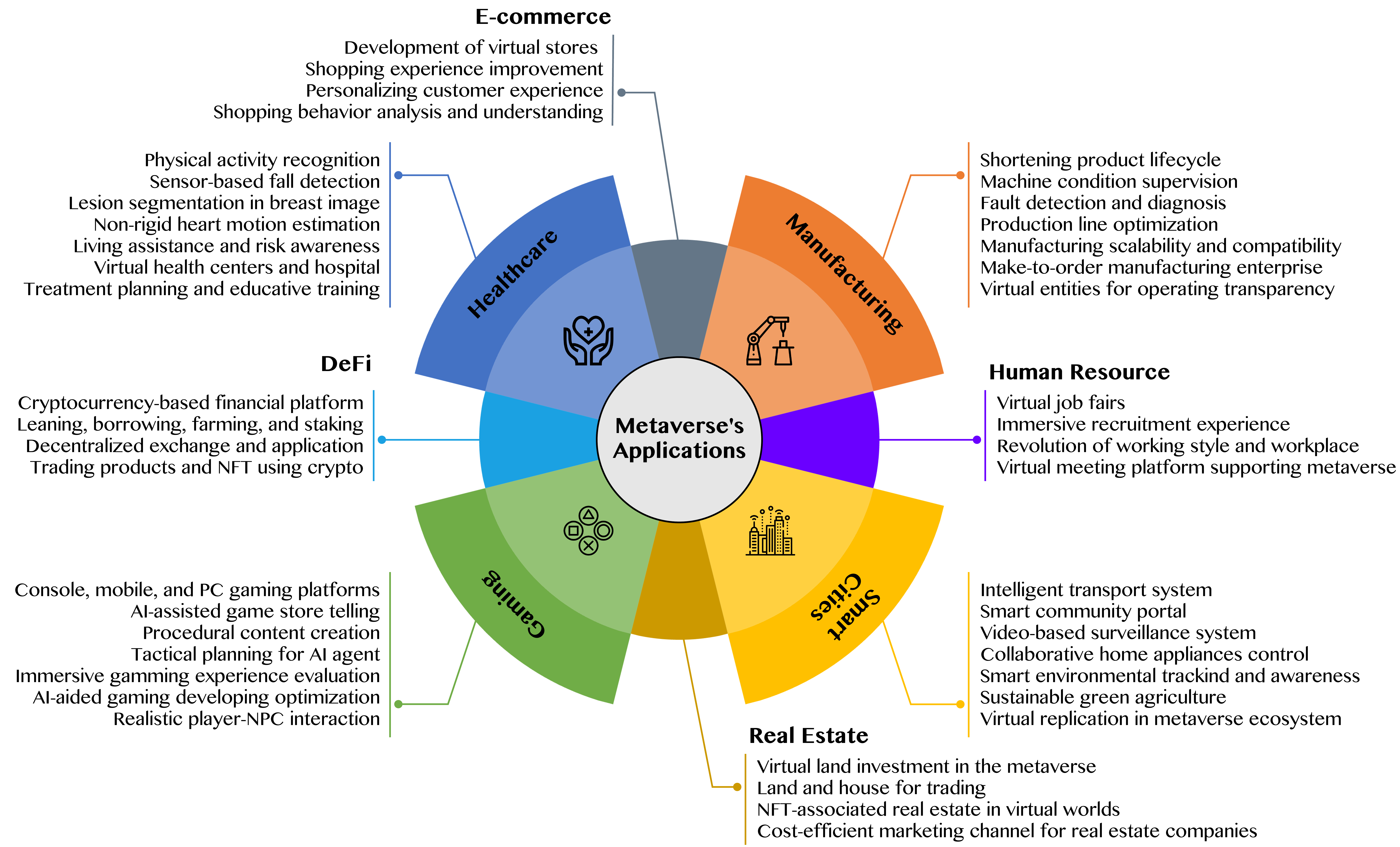}
	\caption{AI for the metaverse in the application aspects with healthcare, manufacturing, smart cities, and gaming besides other promising domains, such as E-commerce, human resources, real estate, and DeFi.}
	\label{fig_applications}
\end{figure*}

This section conveys the existing AI-aided works in the perspectives of four key applications: healthcare, manufacturing, smart cities, and gaming (see Fig.~\ref{fig_applications}); which are probably considered to delivery specialized services in the metaverse 
Besides, some other potential applications, including E-commerce, human resources, real estate, and decentralized finance, are discussed in brief.

\subsection{Healthcare}
The health industry has recently started exploiting some revolutionary techniques like VR and big data incorporated with AI in software and hardware to increase the proficiency of medical devices, reduce the cost of health services, improve healthcare operations, and expand the reach of medical care. 
From a 2D environment to a 3D virtual world, the metaverse allows users to learn, understand, and share patients' health conditions and medical reports in an immersive manner. 
By means of VR/XR systems, AI plays a vital role in many healthcare and medical sectors, for example, achieving better efficiency in providing diagnosis, delivering accurate and faster medical decisions, providing better real-time medical imaging and radiology, and supporting more convenient simulated environments to educate interns and medical students.

In many wearable devices for healthcare and wellness applications and services~\cite{banos2015mining}, AI has been applied to automatically recognize complex patterns of sensory data.
For supporting physicians and health-wellness experts to make decisions in daily living assistance and early healthy risk awareness, a physical activity recognition method was introduced in~\cite{huynh2021physical} by using the sensory data of multiple wearable devices. The method combined the globally handcrafted features and locally deep features (i.e., extracted by a deep CNN) over an intermediate fusion mechanism to improve the activity recognition rate. 
In~\cite{hur2018iss2image}, a novel encoding algorithm, namely Iss2Image, was introduced to convert inertial sensory signal (e.g., accelerometer, gyroscope, and magnetometer) into a color image for CNN-based human activity classification. Furthermore, a lightweight CNN with few layers in a cascade connection was designed to learn the physical activity patterns from encoded activity images.
In~\cite{qian2020wearable}, a system of fall detection using wearable devices was proposed for IoT-based healthcare services, in which a hierarchical DL framework with CNN architectures was developed for collaboratively processing sensory data at local devices and a cloud server. As capably working with multiple wearable devices (e.g., smartphone, smartwatch, and smart insoles), the system yielded high correct detection rate with high data privacy.
Besides CNN, RNN and LSTM networks were exploited to process wearable sensory data in some early healthy risk attentions, such as fall detection and heart failure~\cite{li2020bilstm}.

With the great success of DL, especially CNN architectures, in the image processing and computer vision domains, few recent years have witnessed a vast emergence of DL to address various challenging tasks of medical image analysis because of the requirement of much more specialized knowledge from technicians and medical experts if compared with natural image analysis~\cite{hua2019bimodal,hua2021convolutional}.
For lesion segmentation in breast ultrasound (BUS) images, the work~\cite{ning2021smunet} studied an advanced network, namely saliency-guided morphology-aware U-Net (SMU-Net), by involving an additional middle feature learning stream and an auxiliary network. The coarse-to-fine representative features from the auxiliary network were fused with other features (e.g., background-assisted, shape-aware, edge-aware, and position-aware) to effectively discriminate morphological texture in BUS images.
In~\cite{qi2021nonrigid}, a cost-efficient unsupervised DL approach was introduced to accelerate the processing speed of non-rigid motion estimation of heart in free-breathing 3D coronary magnetic resonance angiography images. Replying on a deep encoder-decoder architecture, the network can learn image similarity and motion smoothness without ground truth information in a patch-wise manner to save computing resources significantly instead of a regular volume-wise manner.
To overcome the obstacle of increasing network size and computation of 3D CNNs in mining complicated patterns in 3D images~\cite{ning2021smunet}, 2D neuroevolutionary networks were investigated for 3D medical image segmentation~\cite{hassanzadeh20212d}, in which an optimal evolutionary 3D CNN was renovated to reduce computational cost without sacrificing accuracy. 
With AI in use as the core technology for data analytics, several healthcare and medical diagnosis applications (e.g., motor rehabilitation and magnetic resonance imaging neurofeedback) can be developed in the VR environment for multipurpose, such as collaborative treatment planning and educative training~\cite{torner2019multipurpose}. 
Indeed, several healthcare and medical services can be provided in the metaverse.
For example, medical students can improve surgical skills by doing interactive practice lessons built for medical education in the virtual world or patients can find some healthcare services via virtual assistants at virtual health centers and hospitals.

\subsection{Manufacturing}
As the current wave of industrial revolution, digital transformation in manufacturing has been happening with digital connection between machines and systems to better analyze and understand the physical entities. Different from digital transformation to enhance the physical world via digital operations, the metaverse creates a virtual world that is translated onto the physical world based on the foundation of reality interaction and persistence. By collaboratively adopting cutting-edge technologies, such as AI and DT, the metaverse for manufacturing can significantly modernize digital operations in the current digital revolution.
Currently, AI with ML algorithms and DL architectures have considerably contributed to the manufacturing domain via numerous industrial applications.

In manufacturing, shortening product lifecycles and increasing the number of product variants are the main reason of high expense for frequent production system reconfigurations and upgrades, especially with ML-based systems which have spent more time and computing resources for new data collection, preprocessing, and model learning. To overcome the above challenges, a symbiotic human-ML framework~\cite{doltsinis2018symbiotic} was leveraged with a reinforcement learning strategy by combining the learning capacity of Q-learning models and the domain knowledge of experts. This framework also considered human exploration to reduce noise in data and improve the quality of automatic decision-making systems.
As a great importance in the modern manufacturing systems, quality inspection has been recently attracting more attention with intelligent data-driven condition supervision approaches; however, they had to face some difficulties from different operating conditions with diversified tasks and applications~\cite{azamfar2020deep}. For reliable fault detection and diagnosis in manufacturing, numerous methods have exploited DL with RNN and CNN architectures to achieve high accuracy while keeping a real-time monitoring. 
For instance, a RNN~\cite{lee2021attention} was developed with an encoder-decoder structure coupled with attention mechanism to predict and diagnose interturn short-circuit faults in permanent magnet synchronous systems. 
In~\cite{xue2020diagnosis}, a data-driven LSTM-based fault diagnosis approach was introduced to early detect multiple open-circuit faults in wind turbine systems. 
In~\cite{guo2019deep}, a DL-based intelligent fault diagnosis method was introduced to address two challenging problems, i.e., the lack of labeled data for learning model and the data distribution discrepancy between training and testing sets, by incorporating CNN architecture and transfer learning mechanism.

Design and implementation of an optimal serial production line are crucial to increase the productivity of the whole manufacturing process. Many recent works have applied AI to optimize some specific sectors in a production system and improve the performance of production line accordingly while meeting scalability and compatibility.
For example, a prediction model~\cite{alkhalefah2021development} was developed to estimate the optimal buffer size in production lines by combining a regular artificial neural network (ANN) and a generic algorithm. The prediction model was further integrated with an optimization mechanism to evaluate and forecast the optimal buffer size in need to maximize productivity. 
In~\cite{huang2019twostage}, an efficient production progress prediction method was formulated with the combination of DL and IoT to optimize the dynamic production and on-time order delivery activities in make-to-order manufacturing enterprises. In the proposed method, a two-stage transfer learning mechanism was executed with historical data and real-time state data using a deep belief network to solve the nonlinearity of production progress.
Nowadays many manufacturing plants have developed industrial collaborative robots to undertake different advanced tasks which require much more cognitive skills, intelligence, and domain knowledge of human to immediately respond unexpected actions or events with high precision and confidence~\cite{lins2021cooperative}.
Therefore, it demands a cooperative AI model to learn complicated patterns from multimodal data for different correlated tasks of manufacturing process and production line, in which the AI model should be equipped with the capability of explanation and reasoning.
Through virtual entities in the metaverse, the industrial manufacturing efficiency is generally improved with AI to speed up production process design, motivate collaborative product development, reduce operation risk to quality control, and obtain high transparency for producer and customers.

\subsection{Smart Cities}

Smart cities acquire the meaningful information about the needs of citizens through the IoT, video cameras, social media, and other sources. Based on the feedbacks automatically collected from users, city governments need to make decisions about which services to remove, offer, and improve. By using more digital tools and pioneering technologies, smart cities will provide smarter interactive services to users over the metaverse platform~\cite{kohli2021review}. The environmental data (e.g., air quality, weather, energy consumption, traffic status, and available parking space) are fully replicated in the virtual world for user-friendly interface. Several smart services, such as utility payment and smart home control, can be now executed in the virtual world via the platforms and systems deployed in the metaverse: intelligent transportation systems (ITS), smart street light management systems, automatic parking systems, smart community portals, and indoor/outdoor video surveillance systems. Currently, the actual impact and benefit of these technologies to the smart cities is limited; however, the metaverse can be an accelerant to spread the presentation of smart services in the daily life of citizens~\cite{mohammadi2018enabling}.

Among different technologies to enable smart cities in the physical world and in the metaverse, AI has shown a great significance to achieve automation and intelligence in smart services. By integrating EEG-based BCI, VR, and IoT technologies fueled by AI, the work in~\cite{park2019development} introduced a steady-state visual evoked potential-based BCI architecture to collaboratively control home appliances. With the visual information captured through head-mounted display, the brain signals were recorded to analyze with ML algorithms and respond control commands with stimulation, which allow users to control home appliances over the IoT network.
In the effort to develop a hybrid ITS from the physical world to its virtual replication in the digital world, the comprehensive work in~\cite{zhu2020parallel} introduced an intelligent and ubiquitous IoT-enabled architecture to control and manage urban traffic. Many scenarios with practical data processing and decision making of different transportation services were investigated simultaneously in both of physical and virtual world, thus conducting high-quality real-time services to users and reducing operation and maintenance costs.
Because of the hastiness of industrialization and the explosion of urbanization, air pollution has become a life-warning problem, which has affected living environment and physical health defectively. 
For early air pollution warning and management, an efficient forecasting approach was studied with a hybrid DL architecture~\cite{du2021deepair}, which combined 1-D CNNs and bi-directional LSTM networks to fully extract intrinsic correlation features and interdependence of multivariate time series data acquired from multiple sensors. 
Besides environmental pollution, sustainable agriculture has been attracting much more concerns in smart green cities~\cite{bhat2021bigdata}. In this context, AI is one of the vital information and communication (ICT) technologies, which has been widely used in precision agriculture systems with yield prediction, quality evaluation, and pest and disease detection. 

 Designing and implementing the metaverse ecosystem for smart cities with all administrative services, such as environment, education, transportation, culture, and other civil services, is really a challenging mission of metropolitan government. By gathering big data from multiple authenticated sources, many administrative services can be provided and improved in the metaverse thanks to AI technology for data analytics, in which usage rules, ethics, and security will be released to guarantee a safe experience environment.

\begin{table*}[ht!]
\setlength{\tabcolsep}{4pt}
\centering
\caption{Summary of AI For The Metaverse in The Application Aspect.}
\begin{tabular}{|p{2.35cm}|p{0.75cm}|p{5.75cm}|p{7.65cm}|}
\hline \hline
\textbf{Application Aspect} & \textbf{Ref} & \textbf{Description} & \textbf{AI Technique} \\ \hline \hline
Healthcare 
& \cite{hur2018iss2image}	& Inertial sensory -based physical activity recognition for healthcare and wellness.	& CNNs with an encoding algorithm to convert inertial sensory signals to color images.\\ \cline{2-4}
& \cite{huynh2021physical}	& Physical activity recognition to support physicians and health-wellness experts in making decisions in living assistance and healthy risk awareness.	& Residual CNNs to extract deep features in combination with handcrafted features based on an intermediate fusion mechanism.\\ \cline{2-4}
& \cite{qian2020wearable}	& Wearable devices based fall detection for IoT-based healthcare and medical services.	& Hierarchical DL framework with CNNs for learning models at local devices and global cloud center.\\ \cline{2-4}
& \cite{ning2021smunet}	& Lesion segmentation in breast ultrasound images for detecting abnormalities.	& Saliency-guided morphology-aware U-Net (SMU-Net) involving an additional middle feature learning stream and an auxiliary network.\\ \cline{2-4}
& \cite{qi2021nonrigid}	& Estimation of non-rigid motion of heart in free-breathing 3D coronary magnetic resonance angiography images.	& Deep encoder-decoder learning framework with CNN architectures.

\\ \hline\hline

Manufacturing
& \cite{doltsinis2018symbiotic}	& Automatic production system reconfigurations and upgrade to shorten product lifecycles and increase the number of product variants.	& Symbiotic with reinforcement learning strategy by  combining the learning capacity of Q-learning models and the domain knowledge of experts.\\ \cline{2-4}
& \cite{lee2021attention}	& \multirow{2}{5.75cm}{Reliable fault detection and diagnosis for quality inspection of production line in manufacturing.}	& Deep encoder-decoder framework using RNN architectures with attention mechanism.\\ \cline{2-2} \cline{4-4}
& \cite{xue2020diagnosis}		& & Data-driven learning model using deep networks with LSTM architecture.\\ \cline{2-2} \cline{4-4}
& \cite{guo2019deep}		& & Data-driven learning model using transfer learning mechanism on CNN architecture.\\ \cline{2-4}
& \cite{alkhalefah2021development}	& Forecasting the optimal buffer size required in production systems to maximize productivity. 	& A combination of a regular artificial neural network and a generic algorithm.\\ \cline{2-4}
& \cite{huang2019twostage}	& Production progress prediction to optimize on-time order delivery activities in in make-to-order manufacturing enterprises.	& Data-driven learning model using two-stage transfer learning mechanism on deep belief network architecture.
\\ \hline\hline

Smart Cities 
& \cite{park2019development}	& Smart home appliances control and management over IoT networks.	& A set of AI algorithms for brain signal processing and decision making to respond stimulation commands.\\ \cline{2-4}
& \cite{zhu2020parallel}	& Intelligent and ubiquitous IoT-enabled urban traffic control and management in ITS systems.	& Hierarchical AI framework by deploying ML algorithms in the digital world to automatically make decisions in the physical world.\\ \cline{2-4}
& \cite{du2021deepair}	& Early air pollution warning and management in smart environment surveillance systems.	& DL framework with 1-D CNNs coupled with bi-directional LSTM networks.\\ \cline{2-4}
& \cite{bhat2021bigdata}	& Sustainable agriculture from smart farm to intelligent food production line.	& AI framework for supervised learning with conventional ML algorithms applied in detection, classification, and recognition models.
\\ \hline\hline

Gaming 
& \cite{synnaeve2016multiscale}	& Modeling multi-scale uncertainty and multi-level abstraction levels in RTS games.	& Multi-scale Bayesian models including Bayesian networks and maps with probabilistic learning for multi-level units control, tactics, and strategy.\\ \cline{2-4}
& \cite{liaqat2020metamorphic}	& Design of a metamorphic testing mechanism for game flow evaluation in artificial chess games.	& ML framework with decision tree algorithm to determine the optimal move among all possible ones for AI agents. \\ \cline{2-4}
& \cite{oh2021creating}	& Development of AI agents in player-vs-NPC and player-vs-player fighting games.	& A combination of RL and deep networks with different architectures, such as RNN and CNN.\\ \cline{2-4}
& \cite{barriga2019improving}	& Development of AI agents and improvement of tactical intelligence in RTS games.	& Supervised learning framework with CNN architectures and reinforcemen learning with deep Q-learning networks. 
\\ \hline\hline

\end{tabular}%
\label{tab_applicationAspect}
\end{table*}

\subsection{Gaming}
Gaming has always been a prime application in the metaverse, in which ML and DL are redefining and revolutionizing the gaming industry across multiple platforms, from console to mobile and PC platforms. This part will explore how ML and DL can revolutionize game development and how building a next gaming generation in the metaverse.

In the last decade, ML has had a huge impact on the way video games are developed. To build more realistic worlds with attractive challenges and unique stories, video game developers and studios have been increasingly turning to ML as a powerful tool set that help systems and NPCs to respond to player’s action dynamically and reasonably.
In~\cite{yannakakis2015panorama}, the role of artificial and computational intelligence in games has been discussed regarding many research topics: NPC behavior strategy and learning, tactical planning, player response modeling, procedural content creation, player-NPC interaction design, general game AI, AI-assisted game story telling, and AI in commercial games. This work further investigated these topics according to three viewpoints: AI algorithms used in each topic, effectiveness of AI to human user in every topic, and human-computer interaction.
For in-game decision making and learning, a comprehensive survey in~\cite{frutos2017review} investigated the use of AI algorithms in intelligent video and computer games. Regarding decision making, some primary AI algorithms (such as decision tree, fuzzy logic, Markov model,  rule-based system, and finite-state machine) were deployed for different game developing tasks: modeling game flow, assessing playing motivation, evaluating immersive experience, adapting gameplay, adapting gaming strategy, customizing gameplay, and modeling and controlling NPC behavior. Besides, many learning-based tasks were accomplished with Naïve Bayes, ANN, SVM, and case-based reasoning system to classify user gameplay, classify NPC behavior, recognize user behavior, and adapt game flow based on personal experience.

In real-time strategy (RTS) games, such as StarCraft, Bayesian models have been used for modeling multi-scale uncertainty and multi-level abstraction levels~\cite{synnaeve2016multiscale}: micromanagement, tactics, and strategy. These probabilistic learning models were able to cope reactive units control, recognize objectives from tactical data, and predict opponent’s gameplay based on strategic information.
To reach an intelligent human-like response mechanism, several game software companies has applied AI in various testing tasks during design and development stages. In~\cite{liaqat2020metamorphic}, a metamorphic testing mechanism was proposed to overcome the impracticability of controlling a large amount of possible moving strategies in artificial chess games. This testing mechanism deployed a decision tree model to reveal metamorphic relations, which in turn effectively determine the optimal move among all possible ones.
With a combination of RL and deep networks, AI agents in~\cite{oh2021creating} were developed to address some inherent difficulties in real-time fighting games and to defeat pro players in one-vs-one battle. In addition to creating different fighting styles through self-play curriculum, this deep RL framework was capable of all two-player competitive games which have level upgrade and balance policies.
RL and supervised learning were also exploited to improve AI agents in RTS games~\cite{barriga2019improving}. Being more superior than the Puppet search algorithm, CNNs and deep Q-learning networks inferred the outcome of costly high-level search and optimized the available time to execute tactical searches. 
In a nutshell, AI with traditional ML and innovative DL algorithms have been making an unprecedented revolution of gaming experience in many aspects: improving the intelligence of NPCs, modeling complex systems, making games more beautiful and rational, conducting more realistic human-NPC interactions, reducing cost of in-game world creation, and opening more opportunities of developing mobile games.
In Table~\ref{tab_applicationAspect}, we summarized the existing application-oriented works utilizing AI technology which have shown the potential to be integrated and deployed in the metaverse.

\subsection{Other Potential Applications}

Besides healthcare, manufacturing, smart cities, and gaming, we have found some auxiliary business applications ahead for the metaverse.

\textit{E-commerce}:
For the purpose of which E-commerce has been integrated into the metaverse, numerous consumer brands have been diving into the digital world to create more delightful and seamless shopping experiences regardless of the unpopularity of VR devices for mainstream consumers. 
Many brands have moved forward step-by-step to build something entirely new by integrating digital stores which are able to bring the best offline and online shopping without any difference in user experience~\cite{elisabeth2021look}.
Indeed, virtual shopping can convey remotely real-time experience of static products, in which consumers, represented by an avatar, can walk around stores in a 3D rendered space and talk with virtual cashier/seller powered by VR and AI technologies.
Personalizing the customer experience is currently attracting more attention from retailers, not only for business survival but also for revenue growth, which can be performed effortlessly in the meta with AI-based shopping behavior understanding.

\textit{Human resources}:
Nowadays, many big tech companies are being creative to seek and communicate with young talents who are looking for jobs. The recruitment manners range from dispatching younger employees/leaders to online interview applicants with video calls to holding job fairs in the metaverse.
Potential applicants can login into the metaverse with blockchain-aided authenticated account and then control their avatars to freely discuss with other avatars representing the company’s human resource managers and project leaders~\cite{lucy2021looking}. For recruitment guidance, the applicants can ask or receive help from the virtual assistant with AI-based NLP. 
In these kinds of recruitment events, the goal is to generate a friendly environment for both the recruiters and applicants for free-style communication, in which the applicants can actively discover more information about job positions instead of passively being asked questions by recruiters.
In the last decade, emerging technologies (such as 5G, IoT, and DL) have brought workers/employees many convenient alternatives (fully remote and hybrid offline-online) to traditional work; however, the metaverse will revolutionize the future of work and the workplace. Recently, Facebook introduced Horizon Workrooms~\cite{Meta2021}, a well-designed meeting platform that allows users, represented as avatars, to work, collaborate, and communicate with others, besides training and coaching activities, in the virtual space by VR devices.

\textit{Real estate}:
We have seen a huge investment from individual investors and institutions to virtual land in the metaverse. Some metaverses have already been released, including virtual gaming platforms like the Sandbox and Axie Infinity and virtual worlds like Decentraland and Upland~\cite{debra2021investors}, in which users can buy, sell, and trade things, including real estate (plots of land and virtual houses). These digital real estates, usually associated by non-fungible tokens (NFTs), are limited by supply to guarantee their values over time based on scarcity. The estate in the metaverse can be used as a virtual place for building constructions (houses and offices) or holding digital events (e.g., art exhibition and fashion show). Moreover, the metaverse is another cost-efficient channel for real estate companies to ultimately show the property to clients before making decision. With VR-aided immersive experience, the clients can discover the property, including interior and exterior from detailed furniture and overall structure, via VR tours and interactive walkthroughs.

\textit{Decentralized finance}:
Based on an open system of finance, decentralized finance (DeFi) is a cryptocurrency-based financial service which is regularly programmed through smart contracts to build exchanges besides providing many major services, such as lending, yield farming, and insurance without centralized authorities. 
Different from centralized finance which is controlled or managed by a centralized entity or a person, DeFi, with blockchain technology, facilitates financial services from peer-to-peer and allows users to fully control their assets while ensuring security and privacy. 
DeFi services are usually delivered via decentralized applications (Dapps) which are entirely built on open-sourced distributed platforms.
By integrating DeFi (including basic and professional services) into the metaverse, users can make purchases virtual products identified by NFTs in the digital world, but will receive the real products in the real life. Furthermore, users can make profits in the metaverse based on the DeFi ecosystem with the lending, borrowing, mining, and staking cryptocurrencies or other tokens. Users can provide liquidity to the liquidity pool with an underlying AI-based mechanism of a decentralized exchange to earn incentives. Swapping tokens (can belong to the same chain or different chains) is the basic service that is prioritized to develop first on any Dapps.

\section{Metaverse Projects}
\label{sec_Project}

\begin{figure*}[!t]
	\centering
	\includegraphics[width=180mm]{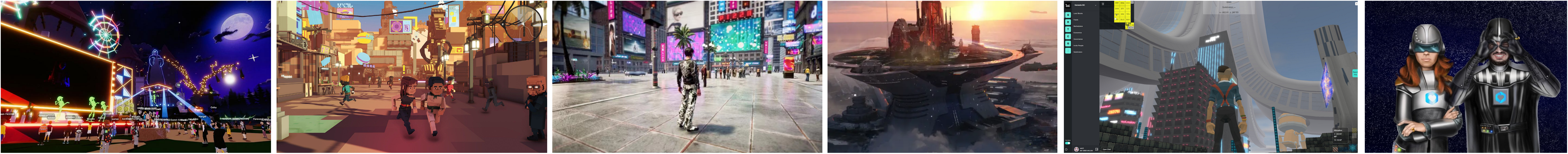}
	\caption{Inside the virtual worlds of different metaverse projects (left to right): Decentraland, Sandbox, Realy, Star Atlas, Bit.Country, and DeHealth.}
	\label{fig_project}
\end{figure*}

This section briefly introduces some attractive metaverse projects, including Decentraland, Sandbox, Realy, Star Atlas, Bit.Country, and DeHealth, which have applied AI to deliver multifarious services and applications in the virtual world, from real estate to E-commerce and real estate. The landscapes inside the virtual worlds of the projects are shown in Fig.~\ref{fig_project}, for DeHealth, virtual doctors as avatars in the metaverse.

\textit{Decentraland\footnote{https://decentraland.org/}}: 
This is a decentralized virtual reality platform built on the Ethereum blockchain, in which users can experience, create, and monetize assets, contents, and applications. In Decentraland, a virtual land is determined as a non-fungible, transferrable, and scare digital asset recorded by the Ethereum smart contract. 
Different from traditional virtual worlds and social networks, Decentraland is not controlled by any centralized organization; that is, no single agent has a permission to modify the rules of software, content, economic of cryptocurrency, or prevent others from accessing the world, trading digital products, and experienced services. 
The traversable 3D world in Decentraland allows embedding immersive component and adjacency to creative content that makes this project attractive and unique. A scripting programming language of Decentraland enables developers to easily code AI-based service-oriented applications for users and accordingly encourage users to create new contents.
Some principal use cases are content curation, advertising, digital collectibles, and social besides other minor ones, such as education, virtual tourism, healthcare, and virtual shopping. Regarding architecture, the Decentraland protocol has three layers: a consensus layer to track land ownership and its content, a land content layer to distribute the materials for rendering via a decentralized storage system, and a real-time layer to establish peer-to-peer connections for world viewing. The native token, for in-world purchase goods and services, of Decentraland is MANA, a fungible token built on the ERC-20 (Ethereum Request for Comments 20) protocol.

\textit{Sandbox\footnote{https://www.sandbox.game/en/}}: 
The Sandbox metaverse is a user-generated decentralized Ethereum blockchain-based virtual world, which allows users and gamers to build, own, and monetize gaming experiences. 
Inspired by Minecraft\footnote{https://www.minecraft.net/en-us}, the Sandbox metaverse is first built as a 2D mobile pixel game and then extended to a fully-fledged 3D world with a voxel gaming platform, wherein users are capable of playing, sharing, collecting, and trading virtual goods and services without central control. Remarkably, creators can earn SAND, the native token of Sandbox, by selling their creations on a marketplace with secure copyright ownership which is associated and guaranteed via NFT, i.e., every item in the metaverse will be authenticated by a unique and immutable blockchain mechanism.
As the primary use cases of SAND in the Sandbox metaverse, the token holders can access and experience the virtual world, vote governance decisions via DAO mechanism, stake tokens to earn revenues, and donate tokens as incentive to developers for the metaverse growth.
Besides the blockchain technology with ERC-20 to generate SAND tokens and ERC-1155 for digital assets trading, AI has been utilized in the Sandbox metaverse. For example, as a powerful toolkit, gaming coders can deploy ML models to improve the intelligence of virtual agents/assistants and DL models to enhance the render quality, and developers can leverage different AI frameworks to minimize gaming crashes and errors. Building and training ML/DL models are trouble-free with intuitive high-level APIs. 

\textit{Realy\footnote{https://realy.pro/}}: 
From the real world to the fully virtual world by creating a unique metaverse ecological world, the Realy metaverse is defined as a super-realistic, futuristic, technologically conscious world, in which E-commerce, social, gaming, and trading are truly integrated to bring a seamless virtual-reality experience to users. In the virtual world of Realy metaverse, users can enjoy colorful journeys through their personalized avatars with 3D virtual clothes which are available in the marketplace and linked with unique NFTs. 
There is an interesting feature of Realy compared with other metaverse projects, that is about the avatar control and management procedures. When users are online, their avatars bring immersive experience. When users are offline, the avatars are driven by a set AI-based self-discipline systems. The whole virtual world is automatically operated, controlled, and managed by a decentralized DAO organization. 
Regarding technology in the Realy metaverse, besides VR and blockchain, AI is adopted in many aspects to generally improve the user immersive experience, such as enhancing 3D visual rendered effects, boosting the intelligence of avatars for realistic behaviors, and integrating VR and holographic projection. 

\textit{Star Atlas\footnote{https://staratlas.com/}}:
As one of the most recent innovative metaverse projects, Star Atlas introduces a virtual gaming world built on the integration of multiplayer video game platforms, real-time immersive experiences with 3D rendered visualization, blockchain-based decentralized financial services, and AI-powered game engines. Star Atlas helps to complete an ecosystem on the Solana blockchain by filling the gap between metaverse and blockchain technology.
In the gaming metaverse of Star Atlas, users can trade digital assets like land, equipment, crew, ship, and components by using the in-game cryptocurrency token known as POLIS which can be used for multiple cross-metaverse games. 
To tune the gameplay more logical and realistic, ML algorithms are applied to improve the intelligence of NPCs and AI agents in tactical planning of actions and fighting strategy in player-NPC combats.

\textit{Bit.Country\footnote{https://bit.country/}}:
As a user-oriented metaverse project, Bit.Country builds a 3D virtual world for everyone who can set up its own community in the metaverse with rules and operations to attract followers and contributors. 
By introducing a new level of virtual social interaction, Bit.Country has two platforms: one is traditional web view for content creation and service provision, and another is addition 3D gaming view for VR-aided immersive experience. 
In each individual community configured by users holding platform tokens, all the rules are managed and linked together by AI models to ensure logical operations without any conflicts. 
Being more than a virtual world, Bit.Country is capable of connecting some virtual aspects to the real world to obtain a sustainable future instead of joyful immersive experience.

\textit{DeHealth\footnote{https://www.dehealth.world/}}:
 Introduced by a British non-profit organization, DeHealth is the world’s first decentralized healthcare metaverse, which allows doctors and patients to work and interact with each other in a full 3D virtual world. In the DeHealth metaverse, several high-quality healthcare and medical services are delivered, such as health analytics on the go, recommendations from advanced AI-bot, and real-time dialogues with doctors and health experts worldwide. To encourage informative data sharing activities, the metaverse has some trading cryptocurrency pools for users and patients to earn assets by selling anonymized medical data. The data collected from a decentralized network will be used to build AI-based diagnostic models for diversified tasks in the healthcare and medical domains. In the metaverse, doctors and patients are able to communicate via a virtual space replicating the real-world environment. Some virtual hospitals and healthcare centers can be built to provide virtual services with real data and diagnosis results. The DeHealth metaverse is in consideration to be extended with additional education-oriented services based on VR technology.

\section{Conclusion and Research Directions}
\label{sec_Conclusion}

In this survey, we have comprehensively investigated the role of AI in the foundation of the metaverse and its potential to enhance user immersive experience in the virtual world. 
At the beginning of this work, the fundamental concepts of the metaverse and AI techniques have been provided, along with the role of AI in the metaverse. 
Subsequently, several principal technical aspects, such as NLP, machine vision, blockchain, networking, DT, and neural interface, and many application aspects, such as healthcare, manufacturing, smart cities, gaming, E-commerce, and DeFi, have been analyzed. The reviewed AI-based solutions have shown that AI has great potential to toughen the systems infrastructure, uplift the 3D immersive experience, and flourish the built-in services in the virtual worlds significantly.
Finally, we have examined prominent metaverse projects, in which AI techniques were used to sharpen the quality of services and encompass the ecosystem of the metaverse.

\begin{figure}[!t]
	\centering
	\includegraphics[width=88mm]{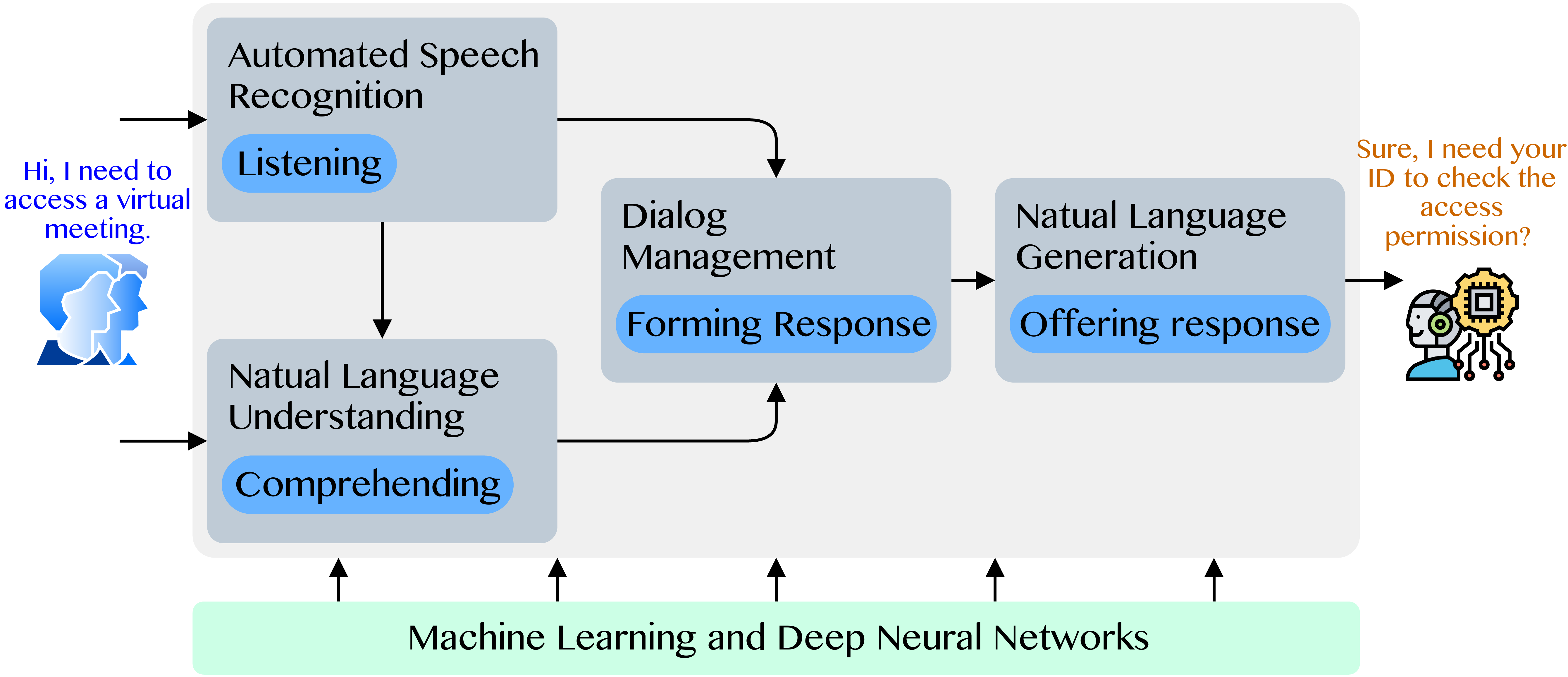}
	\caption{General processing flow of conversational AI to deliver contextual and personal experience to users.}
	\label{fig_conversationalAI}
\end{figure}

We now delineate some AI research directions in the metaverse.
Being more advanced than regular virtual personal assistants which are developed for a general purpose with simple dialog management, virtual customers/employee assistants powered by conversational AI can serve many specific purposes of multi-level philosophical conversations to enhance user interactive experience. 
Conversational AI with a processing flow in Fig.~\ref{fig_conversationalAI} is a set of technologies (e.g., automatic speech recognition, language processing, advanced dialog management, and ML) that can offer human-like interactions in the metaverse based on recognizing speech and text, understanding intention, deciphering various languages, and responding human-mimicking conversations over voice modality. 

Most of the current metaverse projects limit users to explore, own, and customize things in the virtual world. In the future, users will be allowed to create hyperreal objects and content easily and quickly with the help of AI. Various kinds of hyperreal objects (e.g., faces, bodies, plants, animals, vehicles, buildings, and other inanimate objects) can be remixed endlessly by users to make unique experiences and excite creation. Accordingly, the combination of VR and AI-based content generation can bring a complete immersion in alternative realities. In this context, AI tools should be cheap to everyone and have user-friendly interfaces. Further, ethical issues relating to user-generated metaverse need to be seriously examined with constraints and policies between users and third-party organizations to mitigate risks and harmful threats to individuals and societies when users synthesize hyperreal media contents.

In many AI-aided services and applications in the metaverse, the decisions are made by AI agents, which are driven by ML models as black boxes without the capability of interpretability and explainability. Metaverse developers, virtual world designers, and users cannot completely understand AI decision-making processes (e.g., how and why an AI model delivers a prediction), and probably trust them blindly.
To overcome these problems, explainable AI (XAI) is a set of tools and methods to describe AI models, analyze their expected impacts, characterize model transparency, and examine outcomes, allowing human users to entirely comprehend and trust the AI models with end-to-end process monitoring and accountability. 
With XAI, system engineers and data scientists who apply AI in the metaverse (from system infrastructure to services and applications in the virtual worlds) can understand and explain what exactly is happening inside an AI model, how is a specific result generated by an AI algorithm, and when is a prediction model likely to crash.
Besides increasing end user confidence, model auditability, and operative efficiency, XAI mitigates legal risks and security threats of production AI in the metaverse while guaranteeing users' reliable experience.

\balance

\end{document}